\documentclass{ptephy_v1}

\preprintnumber{XXXX-XXXX} 

\usepackage{color}
\usepackage{setspace}
\usepackage{minitoc}

\begin{document}


\title{Measurement of Energy Spectrum of Ultra-High Energy Cosmic Rays}


\author{Valerio Verzi}
\affil{Sezione INFN di Roma \textquotedblleft Tor Vergata\textquotedblright, Roma, Italy \email{valerio.verzi@roma2.infn.it}}

\author{Dmitri Ivanov}
\affil{High Energy Astrophysics Institute and Department of Physics and Astronomy, University of Utah, Salt Lake City, Utah, USA \email{dmiivanov@gmail.com}}


\author{Yoshiki Tsunesada}
\affil{Graduate School of Science, Osaka City University, Osaka, Japan \email{yt@sci.osaka-cu.ac.jp}}

\collaborator{}


\begin{abstract}%

Ultra-High Energy Cosmic Rays (UHECRs) are charged particles of
energies above $10^{18}$ eV that originate outside of the
Galaxy. Because the flux of the UHECRs at Earth is very small, the
only practical way of observing UHECRs is by measuring the {\it
  extensive air showers} (EAS) produced by UHECRs in the atmosphere.
This is done by using air fluorescence detectors and giant arrays of
particle detectors on the ground.  The Pierre Auger Observatory
(Auger) and Telescope Array (TA) are two large cosmic ray experiments
which use such techniques and cover 3000 km$^2$ and 700 km$^2$ areas
on the ground, respectively.  In this paper, we present the UHECR
spectrum reported by the TA, using an exposure of 6300 km$^2$ sr yr
accumulated over 7 years of data taking, and the corresponding result
of Auger, using 10 years of data with a total exposure exceeding 50000
km$^2$ sr yr. We review the astrophysical interpretation of the two
measurements, and discuss their systematic uncertainties.

\end{abstract}


\maketitle

\tableofcontents

\vspace{1cm}

\section{Introduction}
\label{Sec:Introduction}

The origin of cosmic rays (CRs) is an important problem in modern
astrophysics. Extraterrestrial particles of energies greater than
$10^{18}$ and even exceeding $10^{20}$ eV are measured at Earth, while
their arrival directions seem to be distributed almost randomly.  Since the
energies of these CRs extend well beyond the energy range of the solar
particles and energies attainable by the artificial accelerators, the
mechanisms of the CR production and acceleration must be related to
unknown, very energetic phenomena in the universe.

The energy spectrum plays a key role in understanding the
CRs~\cite{RNC}.  The spectrum falls off with energy approximately as a power-law
function, with an average power index of $\gamma \sim 3$, and the CR
flux becomes very small above $10^{14}$ eV.  Above this energy, it is
only practical to detect the CRs indirectly, by measuring secondary
particles, or the {\it extensive air showers} (EAS), which are
produced by the primary CR particles in the atmosphere.

The ultra-high energy CR spectrum has been measured by a number of cosmic ray
experiments~\cite{Verzi-icrc15}, and is known to have 5 features
over 10 orders of magnitude in energy. There is a steepening at $\sim
3 \times 10^{15}$ eV, known as the {\it knee} in the spectrum, a
flattening at $\sim 10^{16}$ eV, and a steepening at $\sim 4 \times
10^{17}$ eV, the so-called the {\it second knee}.  The flattening at
$\sim 10^{16}$ eV is called the {\it low-energy ankle}, a feature
analogous to the widely recognized {\it ankle} at $5 \times 10^{18}$
eV.  At the highest energies, $\sim 5 \times 10^{19}$ eV, there is an
abrupt suppression of the cosmic ray flux to the level of $\sim 1$
particle/km$^2$/century. For a historical overview of observational
studies of CRs and air showers, see e.g. ~\cite{ReviewKH-Watson}.

Although there are indirect evidences from the detection of very high
energy gamma-rays from the supernova remnants that suggest the CRs of
energies below the knee originate within the Galaxy, the
interpretation of CR energy spectrum features remains uncertain.  The
knee and the second knee are believed to be caused by the maximum
acceleration energy available at the Galactic sources, and by the
maximum energies of the magnetic confinement of protons and high-Z
nuclei in the Galaxy.
The gyro-radii of CRs with energies beyond the second knee in the
galactic magnetic field become larger than the size of the Galaxy, and
therefore the magnetic confinement of CRs in the Galaxy is no longer
effective. 
Consequently, the CRs of energies above $10^{18}$ eV, the so-called
Ultra-High Energy Cosmic Rays (UHECRs), must be of extra-galactic
origin.

The UHECRs from distant sources travel tens or hundreds of Mpc before
reaching the Earth, and therefore, interactions and propagation
effects, and the chemical composition must be taken into account in
the interpretation of the CR energy spectrum. Shortly after the
discovery of the cosmic microwave background (CMB) radiation,
Greisen~\cite{Greisen}, Zatsepin, and Kuz'min~\cite{ZK} in 1966
independently predicted a suppression of cosmic ray flux at the
highest energies, as a consequence of photo-pion production from
the interaction of the CRs with the low energy CMB photons.  This
feature is now called the {\it GZK cutoff}.
In the case of a pure proton composition, the {\it ankle} can be
explained by the electron-positron pair-production from 
interactions of the CR protons with the CMB
photons~\cite{Berezinsky1}.  In the case of a mixed composition, on
the other hand, propagation effects are complicated by the fact
that the primary nuclei also suffer interactions that cause
a progressive reduction of their mass numbers. Other
alternative models assert that a cut-off on the acceleration mechanism
at the sources may play some role in the explanation of the observed
suppression of the cosmic ray flux~\cite{Disappointing}.

\begin{figure}[h]
\centering
\includegraphics[width=0.8\textwidth]{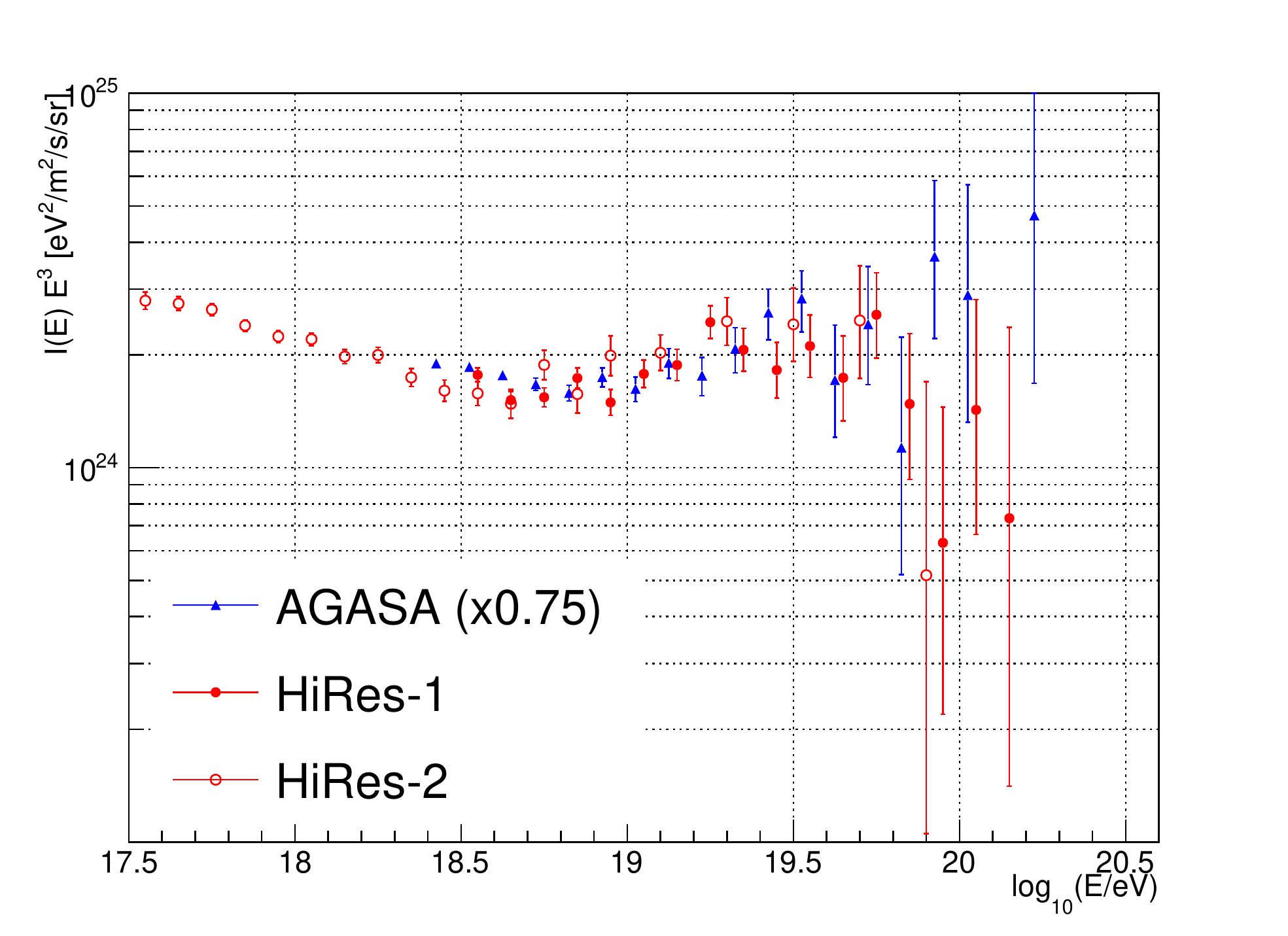}  
\caption{Energy spectra measured by the AGASA and HiRes. The energy of
  AGASA data points is scaled down to $75\%$ to match the position of
  the ankle with HiRes at $\sim 5 \times 10^{18}$ eV.}
\label{Fig:AGASA-HiRes}
\end{figure}

Historically, observation of the cut-off in the energy spectrum was a
technically challenging task. Because the rate of CRs of energies
greater than $10^{20}$ eV is as low as $1$ event per square kilometer
per century, experiments with very large effective areas, long
observation periods, and good energy resolution were required to see
the effect.  AGASA (Akeno Giant Air Shower Array)~\cite{AGASA} and
Hi-Resolution Fly's Eye (HiRes)~\cite{HiRes} were the first cosmic ray
detectors large enough to measure the energy spectrum of UHECRs above
10$^{19.0}$ eV, as shown in Fig. \ref{Fig:AGASA-HiRes}. There were two
major differences in their results. First, there was a difference in
the overall energy scale, which came from the difference in the
techniques employed by the two experiments. AGASA used an array of
scintillation counters that were detecting EAS particles at the ground
level, while HiRes employed fluorescence detectors that were sensitive
to fluorescence light emitted due to the energy deposition of the EAS
particles in the atmosphere.  The systematic uncertainties in
determining the CR primary energy were $\sim~20~\%$ in both
experiments.  The second important difference was in the shape of the
spectrum above the {\it ankle}. The HiRes spectrum showed a steepening in
the spectrum at $6 \times 10^{19}$ eV ~\cite{HiRes} as predicted by
the GZK theory ~\cite{Greisen,ZK}, whereas the AGASA spectrum extended
well beyond the cut-off energy ~\cite{AGASA}.  The tension between the
two major experiments in the 1990's led to an idea of the {\it hybrid}
detection of UHECRs, where both surface detectors and fluorescence
detectors can be used within a single experiment.  The Telescope Array
(TA)~\cite{TA-icrc15-HL} and Pierre Auger Observatory
(Auger)~\cite{Auger-icrc15-HL} are modern hybrid cosmic ray
experiments.

This paper describes recent measurements of the UHECR spectrum by the
Telescope Array (TA) experiment ~\cite{TA-icrc15-HL} and the Pierre
Auger Observatory (Auger)~\cite{Auger-icrc15-HL}.  The TA is a cosmic
ray observatory that covers an area of about 700 km$^2$ in the
northern hemisphere, and Auger has an effective area of 3000 km$^2$ in
the southern hemisphere.  Both experiments use two types of
instruments, surface detectors (SDs) and fluorescence detectors
(FDs).  The {\it hybrid} detection technique, where the CR showers are
simultaneously observed with the FDs and SDs at the same site, 
allows a very precise
determination of the CR energies and arrival directions.  The FDs
measure fluorescence light emitted by the atmospheric molecules
excited by the charged particles in the EAS, and observe the
longitudinal development of the EAS using mirror telescopes coupled
with clusters of photo-multiplier tubes.  Because the FDs are mostly
sensitive to the calorimetric energy deposition in the atmosphere, the
energy determination of the primary CRs is nearly independent of the
details of the hadronic interactions within the EAS, where there are
considerable uncertainties in different models.
The FDs operate
at a $\sim 10\%$ duty cycle because the FD data can be collected only
during nights with low moonlight background and with dry air and clear skies. The SDs, on the
other hand, directly measure EAS particles at the ground level at a nearly
$100\%$ duty cycle, regardless of the weather conditions.

This paper is organized as follows.  The TA and Auger detectors are
described in Sec.~\ref{Sec:TAAuger}. 
The details of the energy scale in the two experiments are summarized and discussed in
Sec.~\ref{Sec:EnergyScale}.  The latest results of the energy spectrum
measurements and their astrophysical interpretations are given in
Sec.~\ref{Sec:Spectrum} and~\ref{Sec:Discussion}. 
Sec.~\ref{Sec:Outlook} is devoted to the
discussion on the current status and future prospects of the UHECR
field, as well as the existing plans of extension and upgrade of the
detectors foreseen by the TA and Auger collaborations.

\section{Telescope Array and Auger Instruments}
\label{Sec:TAAuger}

\subsection{TA Detectors}
\label{Sec:TAdetectors}
\begin{figure}[h]
\centering
\includegraphics[width=1.0\textwidth]{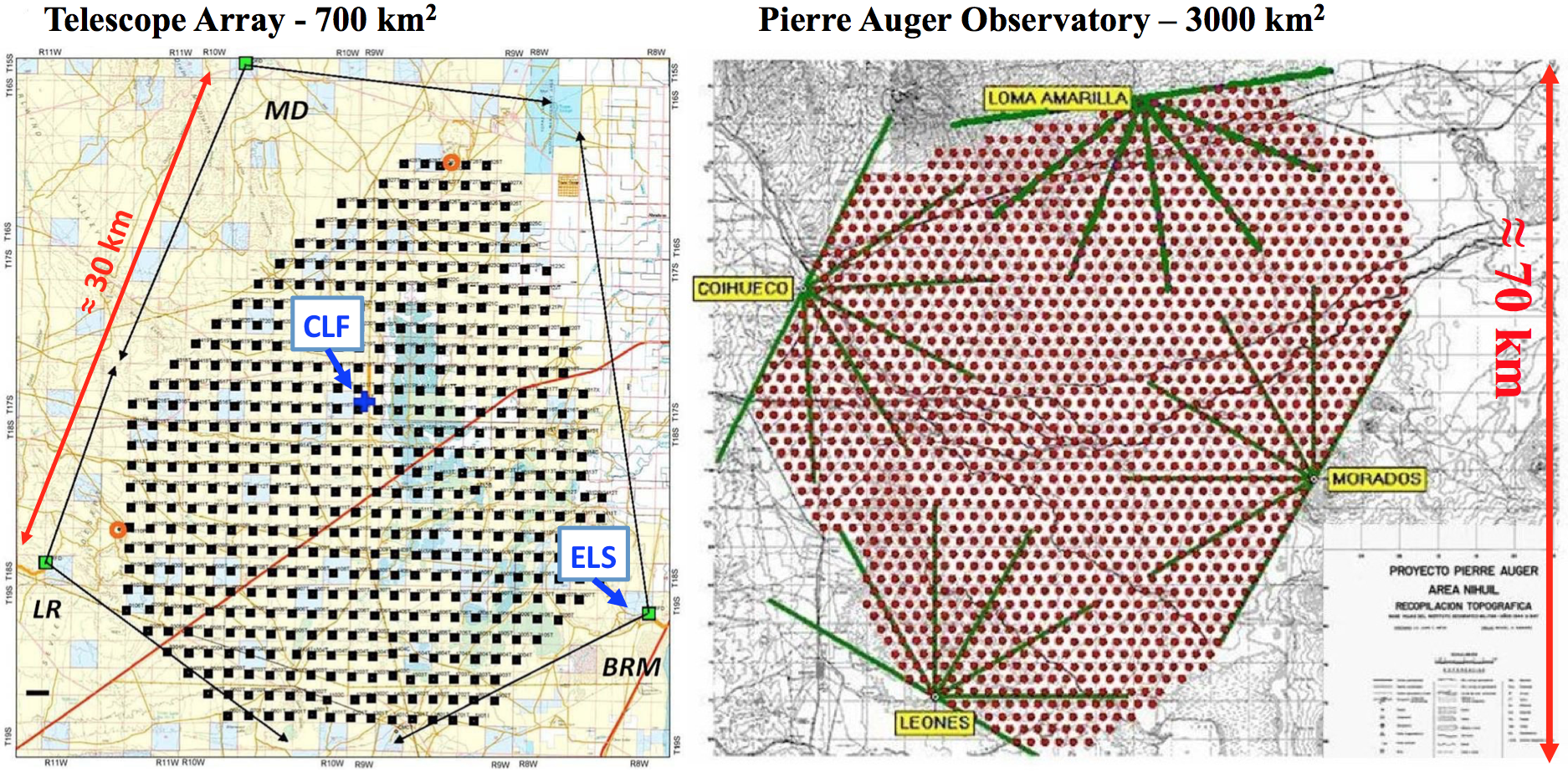}  
\caption{The Telescope Array experiment (left) and Pierre Auger
  Observatory (right). Surface detector units are represented by the
  dots and the FDs are located on the perimeters of the sites.}
\label{Fig:Site}
\end{figure}

The Telescope Array experiment~\cite{TA} is located in Millard County,
Utah (USA) at a 39.3$^\circ$~N latitude and $\sim$~1400~m altitude
above sea level.  
The TA detectors have been in full operation since May
2008.  A sketch of the TA site is shown in the left panel of Fig.~\ref{Fig:Site}.
 
\begin{figure}[h]
\centerline{ \includegraphics[width=0.34\textwidth]{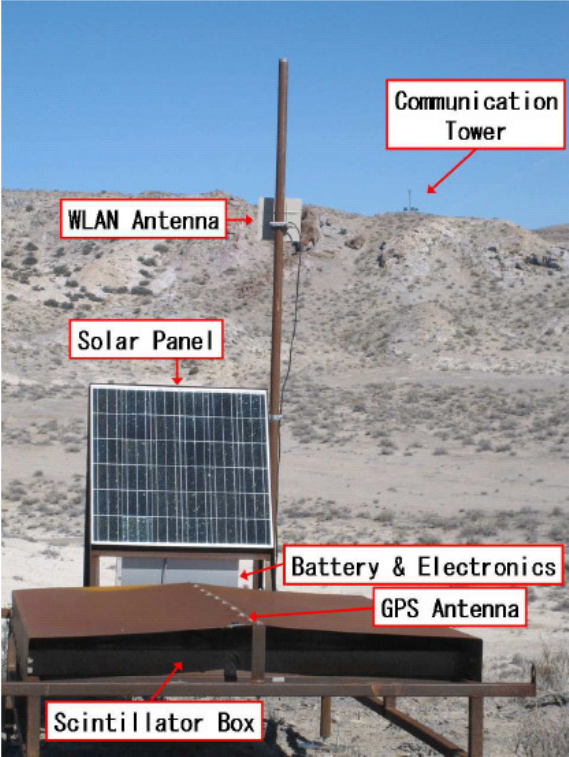} \hspace{1cm} \includegraphics[width=0.38\textwidth]{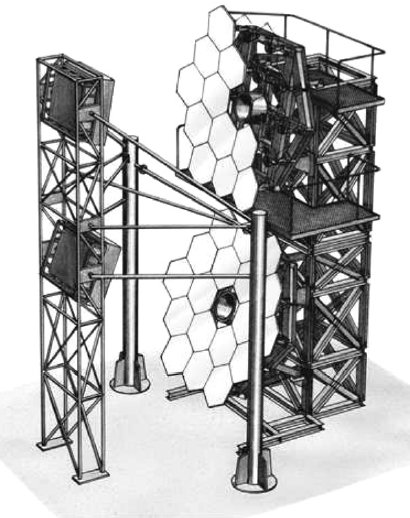}   }
\caption{TA SD unit (left) and FD telescope (right).}
\label{Fig:TAInstruments}
\end{figure}

The TA SD consists of 507 particle counters arranged on a 1.2~km spaced square
grid and covers an area of $\sim$700~km$^2$ on the ground.  Each
surface detector unit, shown on the left panel of
Fig.~\ref{Fig:TAInstruments}, consists of two layers of 
3~${\rm m^{2}}$, 1.2~cm thick plastic scintillators~\cite{SD-TA}.
Scintillation light in each layer is collected and directed to the
photo-multiplier tube (PMT) by the wavelength-shifting fibers.
There is one PMT for each layer. Outputs of the PMTs of the upper and
the lower layers are individually digitized by 12 bit flash
analog-to-digital converters (FADCs) at a 50 MHz sampling rate.

The TA includes three fluorescence detector stations that 
overlook the surface array. Middle Drum (MD) FD is located in the
northern part of the TA, and Black Rock Mesa (BR) and Long Ridge (LR)
FDs are in the southern part.

The MD station utilizes 14 refurbished telescopes previously used in
the High-Resolution Fly's Eye (HiRes) experiment~\cite{FD-TA-MD}. Each
telescope consists of a $\sim~{\rm 5~m^{2}}$ spherical mirror and an imaging
camera of 256 PMTs that uses a sample-and-hold readout system. The
telescopes are logically arranged in two layers, called {\it rings},
that observe two different elevations. Physically, the MD telescopes
are arranged in pairs: ring 2 telescopes that observe higher
elevations are placed next to the ring 1 telescopes. The station
covers 110$^\circ$ in azimuth and 3$^\circ$ to 31$^\circ$ in
elevation.

Black Rock Mesa and Long Ridge stations have each 12 fluorescence
telescopes that are also arranged in two rings.  The telescopes use a
new design shown in the right panel of
Fig.~\ref{Fig:TAInstruments}. Each mirror of the BR and LR telescopes
is composed of 18 hexagonal segments. The radius of curvature of each
segment is $6067$~mm, and the total effective area of the mirror is
$6.8~{\rm m^2}$. The imaging camera of a BR-LR telescope consists of
256 PMTs that are read out by a 40 MHz FADC system. Each station covers
108$^\circ$ in azimuth and 3$^\circ$ to 33$^\circ$ in
elevation~\cite{FD-TA-BRM-LR}.

The calibration of the TA FDs was carried out by measuring the {\it
  absolute} gains of dozens of \textquotedblleft standard\textquotedblright ~PMTs that are installed in
each camera using the {\it CRAYS} (Calibration using Rayleigh
Scattering) system in the laboratory~\cite{FD-TA-AbsCal}.  The rest of
the PMTs in the cameras are {\it relatively} calibrated to the
standard PMTs by using Xe lamps installed at the center of each
mirror~\cite{FD-TA-RelCal1,FD-TA-RelCal2}.

\subsection{Auger Detectors}
\label{Sec:Augerdetectors}
The Pierre Auger Observatory~\cite{Auger} is located in a region
called {\it Pampa Amarilla}, near the small town of Malarg\"ue in the
province of Mendoza (Argentina), at $\sim$35$^\circ$~S latitude and an at 
altitude of 1400 m above sea level. 
The observatory is in operation since 2004 and its construction was completed 
in 2008.
A sketch of the site is shown in
the right panel of Fig.~\ref{Fig:Site}.

\begin{figure}[h]
\centerline{ \includegraphics[width=0.47\textwidth]{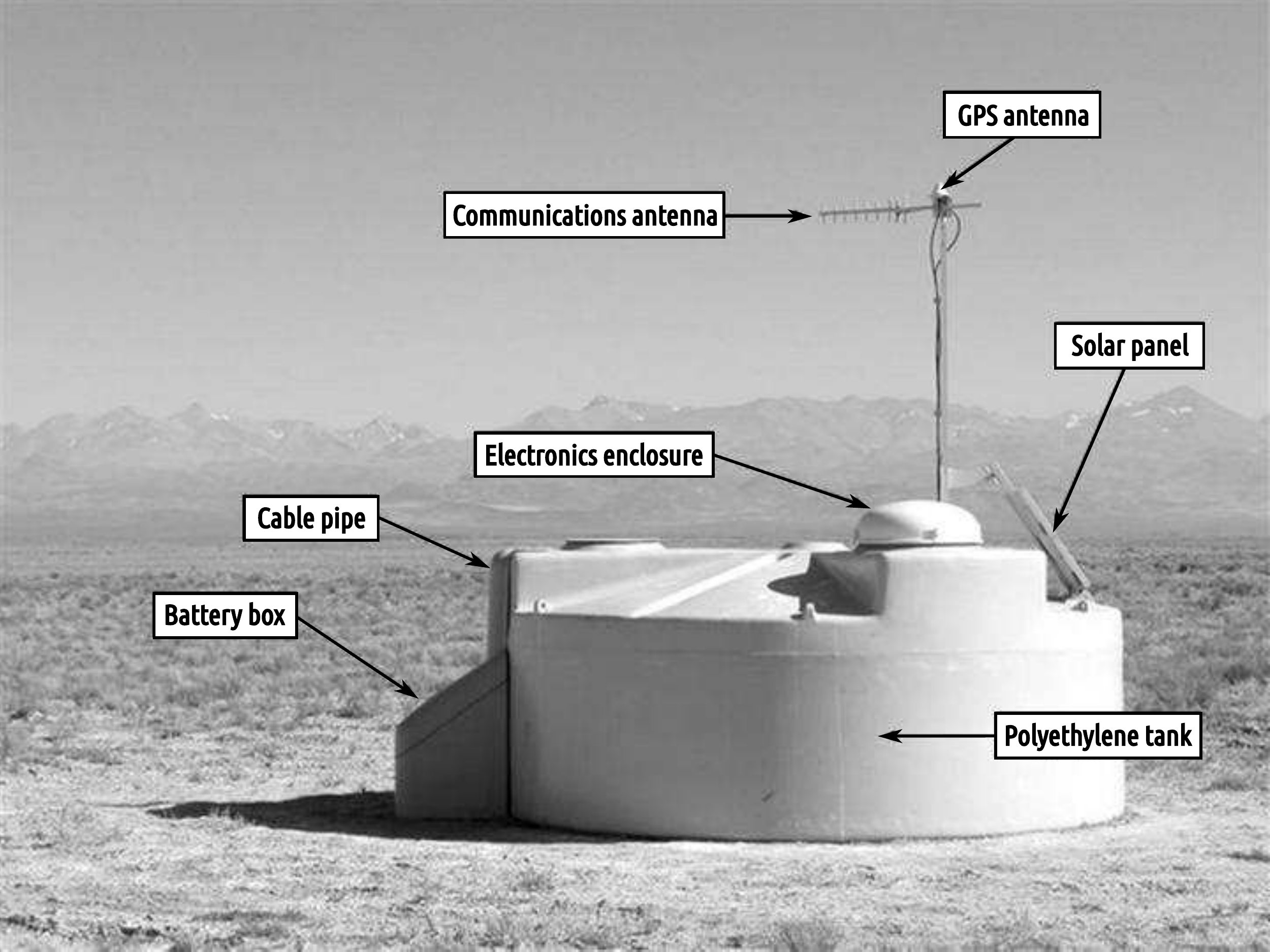} \hfill \includegraphics[width=0.45\textwidth]{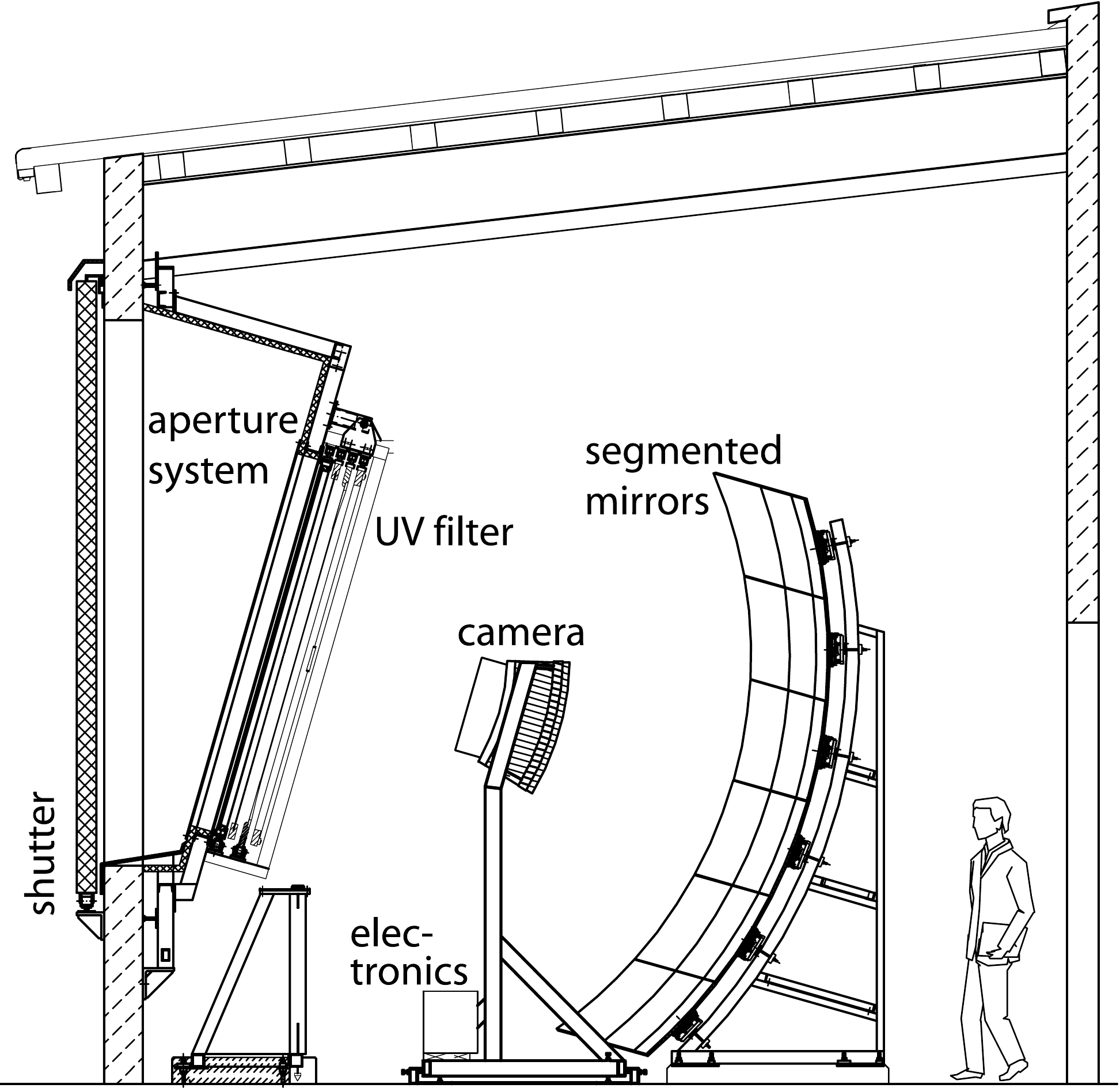}   }
\caption{Auger SD unit (left) and FD telescope (right).}
\label{Fig:AugerInstruments}
\end{figure}

The Auger SD consists of 1660 water-Cherenkov detectors (WCDs)
arranged on a hexagonal grid of 1.5~km spacing. The effective area of
the array is $\sim$3000 ${\rm km}^2$. Each WCD unit, shown in the left
panel of Fig.~\ref{Fig:AugerInstruments}, is a plastic tank of a
cylindrical shape, 10~m$^2$ $\times$ 1.2~m, filled with purified
water.  Cherenkov radiation, produced by the passage of charged
particles through the water, is detected by the three PMTs, 9'' in
diameter each. The signal of the PMTs is digitized by an FADCs at a 40
MHz sampling rate.  Because WCDs extend $1.2$~m in the vertical
direction, the Auger SD is sensitive to the cosmic ray showers
that are developing at large zenith angles.

The Auger FD consists of 24 telescopes placed in four buildings
located along the perimeter of the site. A sketch of a telescope is
shown in the right panel of Fig.\ref{Fig:AugerInstruments}.  Each
telescope has a 3.5 m $\times$ 3.5 m spherical mirror with a curvature
radius of 3.4 m.  The coma aberration is eliminated using a Schmidt
optics device, which consists of a circular diaphragm of radius 1.10 m
and a
series of corrector elements mounted in the outer part of the aperture. 
An ultraviolet transmitting filter is
placed at the telescope entrance in order to reduce the background
light and to provide the protection from the outside dust.  The focal
surface is covered by 440 PMTs, 22 rows x 20 columns, and the overall
field of view of the telescope is $30^\circ$ in elevation and
$28.6^\circ$ in azimuth. The PMTs use photocathodes of an hexagonal
shape and are surrounded by light concentrators in order to
maximize the light collection and to guarantee a smooth transition
between the adjacent pixels.  The signal from each PMT is digitized by
a 10 MHz FADC with a 12 bit resolution.

The Auger FDs are calibrated using a portable cylindrical diffuser,
called the {\it drum}~\cite{Auger-Drum}.

During the calibration process, the drum is
mounted to the aperture of each telescope, and provides a uniform
illumination of the entire surface area that is covered by the PMTs of
the telescope.  The drum is absolutely calibrated using a
NIST-calibrated photodiode, and provides an absolute end-to-end
calibration of all pixels and optical elements of every Auger FD
telescope.  The long-term time variations in the calibration of the
telescopes are monitored using LED light sources that are installed in
each building.

\subsection{Auxiliary Facilities at the TA and Auger}
\label{Sec:AuxiliaryFacilities}

In order to reconstruct the shower energy from the FD information
accurately, it is necessary to know the attenuation of the light due
to the molecular and aerosol scattering as the light propagates from
the shower to the detector.  The molecular scattering can be
calculated from the knowledge of the air density as a function of
height, and the aerosol content of the atmosphere is monitored each
night during the FD data collection.  The aerosols are measured in the
TA and Auger using similar instruments.  These include central
laser facilities (CLFs) placed in the middle of the arrays, and 
standard LIDAR (LIght Detection and Ranging)
stations~\cite{TALIDAR,Auger}. Auger CLF has recently been
upgraded to include a backscatter Raman LIDAR receiver. Other
instruments like infra-red cameras are also employed in both
experiments to continuously monitor the cloud coverage.

Both the TA and Auger collaborations have enhanced the baseline
configurations of their detectors to lower the minimum detectable
shower energies.  The TA low energy extension (TALE) consists of 10
additional fluorescence telescopes viewing higher elevation
angles~\cite{TA-icrc15-TALE1,TA-icrc15-TALE2}, from $32^\circ$ to
$59^\circ$, installed at the MD site, and an infill array of the same
scintillation detectors as those used by the main TA SD array. In a
similar way, Auger has installed three additional High Elevation Auger
Telescopes (HEAT), viewing from 30$^\circ$ to 60$^\circ$ in elevation,
at the Coihueco site (see Fig.~\ref{Fig:Site}). HEAT overlooks a 27
km$^2$ region on the ground that is filled with additional WCDs using
750 m spacing~\cite{Auger}.

An important calibration facility, called the {\it Electron Light
  Source} (ELS)~\cite{TA-ELS}, has been implemented by the TA
collaboration. The ELS is a linear accelerator installed in front of
the TA Black Rock Mesa FD station at a distance of 100~m from the
detector.  The ELS provides a pulsed beam of 40~MeV electrons that are
injected into the FD field of view.  The pulse frequency is 1~Hz and
each pulse has a duration of 1~${\rm \mu s}$ and an intensity of about
$10^{9}$ electrons.  The ELS beam mimics the cosmic ray air showers
and provides an effective test not only for the FD calibration but
also for the other kinds of detectors, such as radio antennas.

\section{TA and Auger Energy Scale}
\label{Sec:EnergyScale}
Both TA and Auger experiments use FD measurements to set their energy
scale. The FD measures fluorescence photons produced by de-excitation
of the atmospheric molecules (nitrogen and oxygen) that have been
excited by the charged particles in the EAS, and provides a nearly
calorimetric estimate of the total EAS energy.  The fluorescence
photons are emitted isotropically in the wavelength range between 290
and 430 nm.  The most significant line emission at 337 nm contributes
$\sim 25\%$ of the total emission intensity.  The number of emitted
photons is proportional to the energy deposited by the charged
particles in the EAS.  The proportionality factor, called the {\it
  fluorescence yield} (FY), is measured by several experiments using
accelerator beam and radioactive sources (see~\cite{ArquerosLast} for
a review on this topic).

\begin{figure}[h]
\vspace{0cm}
\centering
\centerline{ \includegraphics[width=0.51\textwidth]{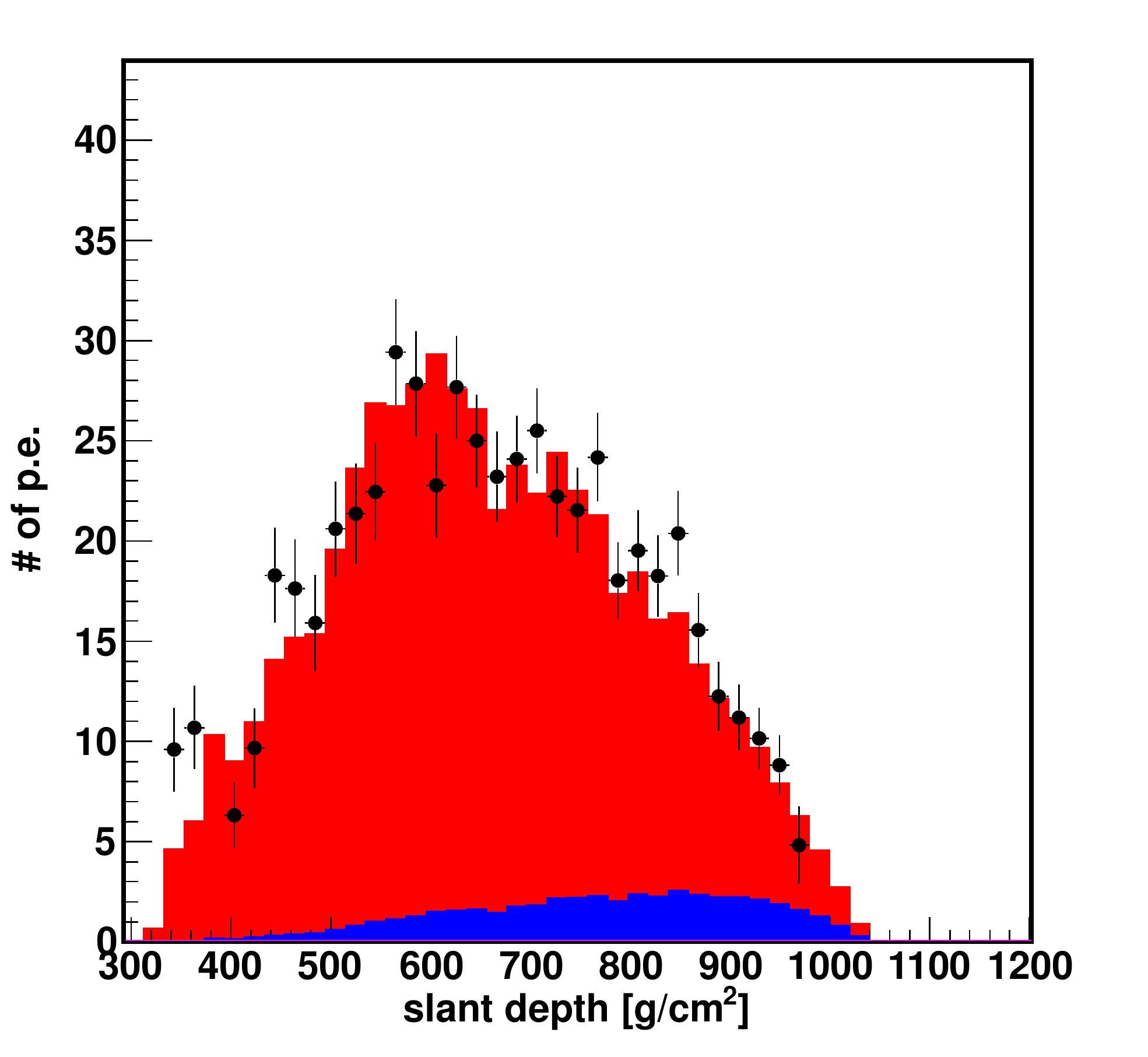}  \includegraphics[width=0.47\textwidth]{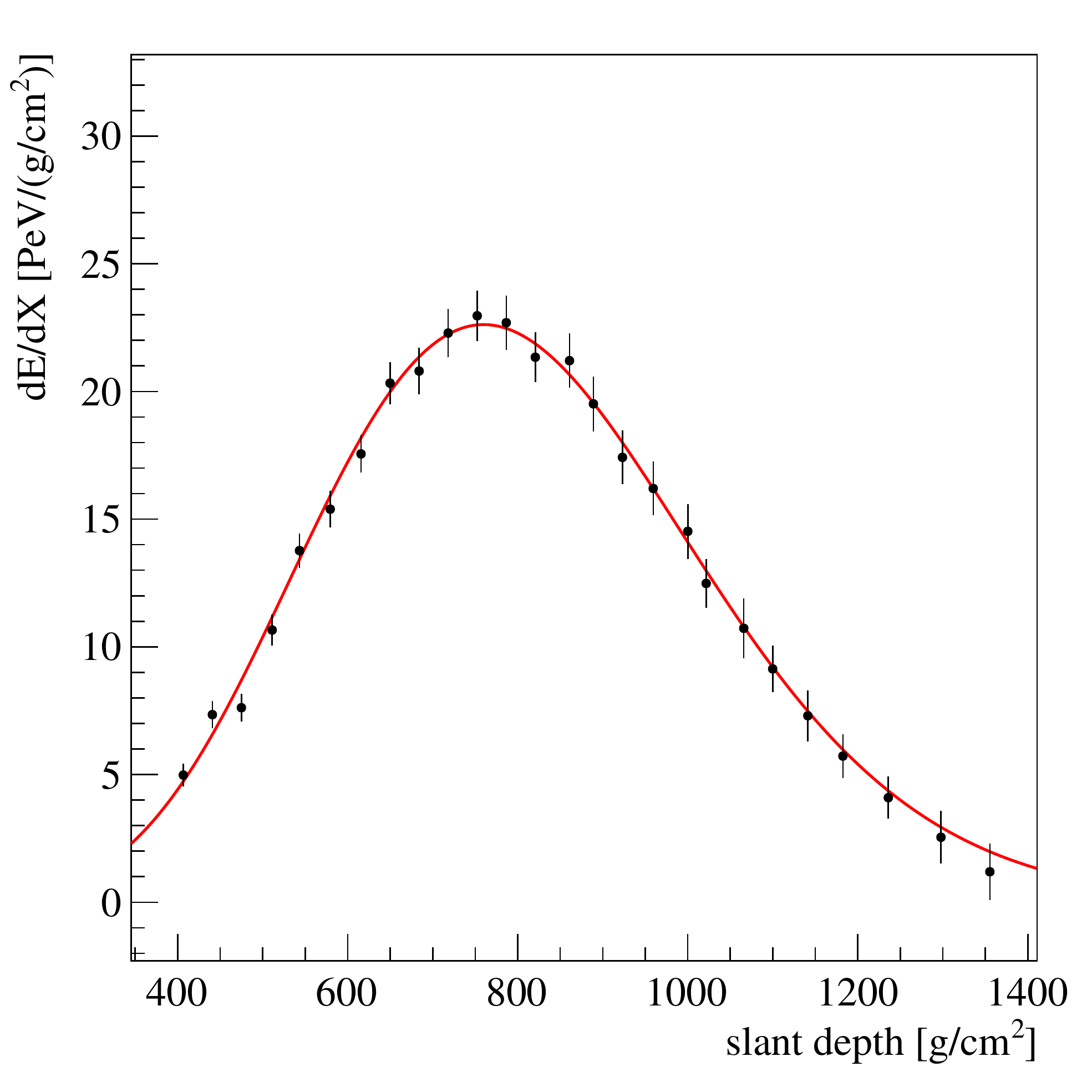}   }
\vspace{0cm}
\caption{Shower longitudinal profiles, as seen by the TA and
  Auger. The TA profile (left) shows the numbers of detected
  photoelectrons versus slant depth along the shower propagation axis.
  Points with error bars represent the data, reconstructed
  fluorescence light is shown in red color, and the Cherenkov
  contribution is shown in blue~\cite{TA-HybEnSc}. In the case of
  Auger (right), reconstructed energy deposition (points with error
  bars) is plotted versus slant depth along the shower axis. Red curve
  shows the Gaisser-Hillas function fitted to the data points~\cite{Unger-FDRec}.} 
\label{Fig:FDProfile}
\end{figure}


As the EAS develops in the atmosphere, fluorescence light emitted at
different altitudes triggers the FD pixels (PMTs) at different times.
Pointing directions of the triggered pixels and the pixel time
information are used to reconstruct the full geometry of the cosmic
ray shower event, which includes the event arrival direction and the
impact parameter (distance between the FD station and the EAS axis).
Additionally, information from the triggered surface detector stations
on the ground can be added to constrain the FD timing fit and improve
the resolution of the EAS geometry.  Events that are reconstructed
using FD and SD information simultaneously are called the {\it hybrid
  events}.

Examples of reconstructed FD longitudinal profiles are shown in
Fig.~\ref{Fig:FDProfile}.  The energy deposition $(dE/dX)$ is
determined as a function of the slant depth $X$ along the shower axis
using the intensity of the signal of the triggered pixels. The $dE/dX$
reconstruction procedure requires the absolute calibration of the FD
telescopes, knowledge of the attenuation of the light due to the
scattering by the air molecules and the aerosols, and the absolute
fluorescence yield.  The integral of the $dE/dX$ profile gives the
calorimetric energy of the shower: 
$E_{\rm cal} = \int \left(dE/dX\right) dX $. The
total (and final) energy of the primary cosmic ray is obtained from
$E_{\rm cal}$ after the addition of the so called {\it invisible}
(or equivalently {\it missing}) energy $E_{\rm inv}$, which is the
energy that is carried away by the high-energy muons and neutrinos
that do not deposit their energies in the atmosphere and thus cannot
be seen by the FD.  For a typical EAS, $E_{\rm inv}$ is of the order
of 10 to 15\% of the total primary energy. Further details on the
energy scale and {\it invisible} energy corrections in TA and Auger
will be discussed in Sec. \ref{SubSec:SystUnc} and
\ref{Sec:EnergyScaleComparison}.

\subsection{Surface Detector Energy Reconstruction}
\label{SubSec:SDCalib}

The SD energy scale in both TA and Auger is calibrated by the FDs
using well reconstructed hybrid events.  This is done by comparing the
SD energy estimators with the energies obtained from the corresponding
FD longitudinal profiles on an event by event basis.

The SD energy estimators in TA and Auger are obtained using
conceptually similar analyses~\cite{TA-EnSp-Comb,Auger-EnSp}.  The
energy of the primary CR particle, arriving at a given fixed zenith
angle $\theta$, is assumed to be directly related to the intensity of
the shower front at a certain distance from the shower core.  This
relation depends on $\theta$ because the effective amount of the air
material that the EAS propagates through, before reaching the ground
level, increases as $1/{\rm cos}(\theta)$.

\begin{figure}[h]
\vspace{0cm}
\centering
\centerline{ \includegraphics[width=0.41\textwidth]{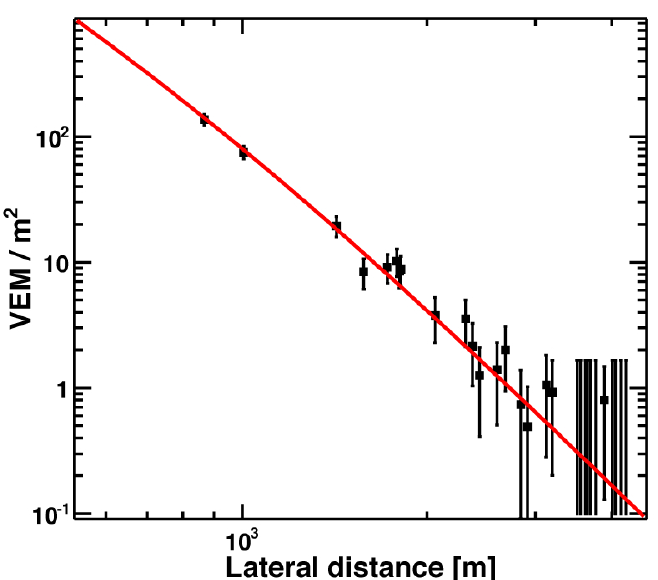}  \includegraphics[width=0.5\textwidth]{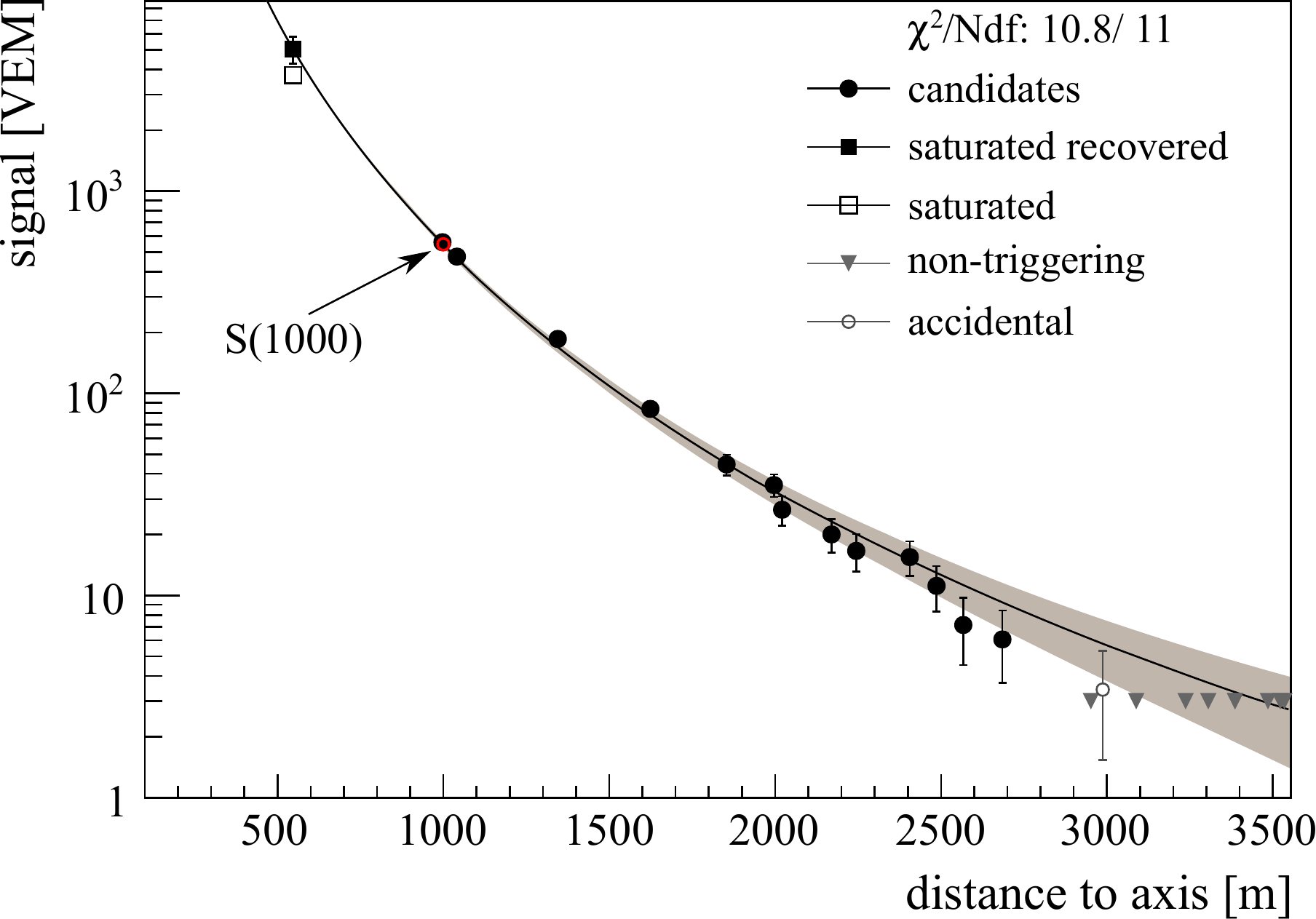}   }
\vspace{0cm} 
\caption{Lateral distribution of the air shower energy deposition
  measured by the SD for typical TA~\cite{DmitriThesis} (left) and
  Auger~\cite{Auger} (right) events. Black points represent the
  individual surface detector stations and solid lines represent the
  fits to the analytic curves. }
\label{Fig:SDRec}
\end{figure}

The shower axis and the point of impact on the ground are determined
from the timing and the intensity of the signal in the triggered SD
stations.  The best energy estimator is obtained by evaluating the
intensity of the shower front at an optimum distance $r_{\rm opt}$
from the shower core. This is done using analytic functions with
parameters determined from the fit to the intensity of the signal as a
function of the distance from the shower core (see
Fig.~\ref{Fig:SDRec}).  The optimal energy determination distance for
the TA SD with 1200 m spacing is $r_{\rm opt} = 800$ m.  
For Auger, whose spacing is 1500 m, $r_{\rm opt}$ is 1000 m. 
A similar reconstruction technique is used for the Auger events detected by the 750 m 
array placed in front of the HEAT telescopes (see Sec.~\ref{Sec:AuxiliaryFacilities}). In this 
case $r_{\rm opt}$ is 450 m.

Next, the intensity of the signal at the optimal distance $r_{opt}$ is
corrected for the zenith angle attenuation. In TA, this correction is
made using a detailed Monte Carlo
simulation~\cite{TA-SDsp2013,DmitriThesis}, and an energy estimator
$E_{\rm SDMC}$ is obtained. $E_{\rm SDMC}$ represents the reconstructed TA SD
energy prior to the calibration of the energy scale by the
FD. Standard TA SD reconstruction uses events with $\theta <
45^\circ$.

In the case of Auger, the zenith angle attenuation is derived from the
data using the \textquotedblleft Constant Integral Intensity
Cut\textquotedblright ~method~\cite{CIC}. The resulting Auger energy
estimators are called $S_{38}$ for the 1500 m array and $S_{35}$ for
the 750 m and array, and represent the signal values the EAS would
have produced if it arrived at the zenith angles of $38^\circ$ and
$35^\circ$, respectively.  These numbers correspond to the median
values of the event zenith angle distributions of the Auger 1500 m and
750 m arrays.

\begin{figure}[h]
\vspace{0cm}
\centering
\centerline{ \includegraphics[width=0.5\textwidth]{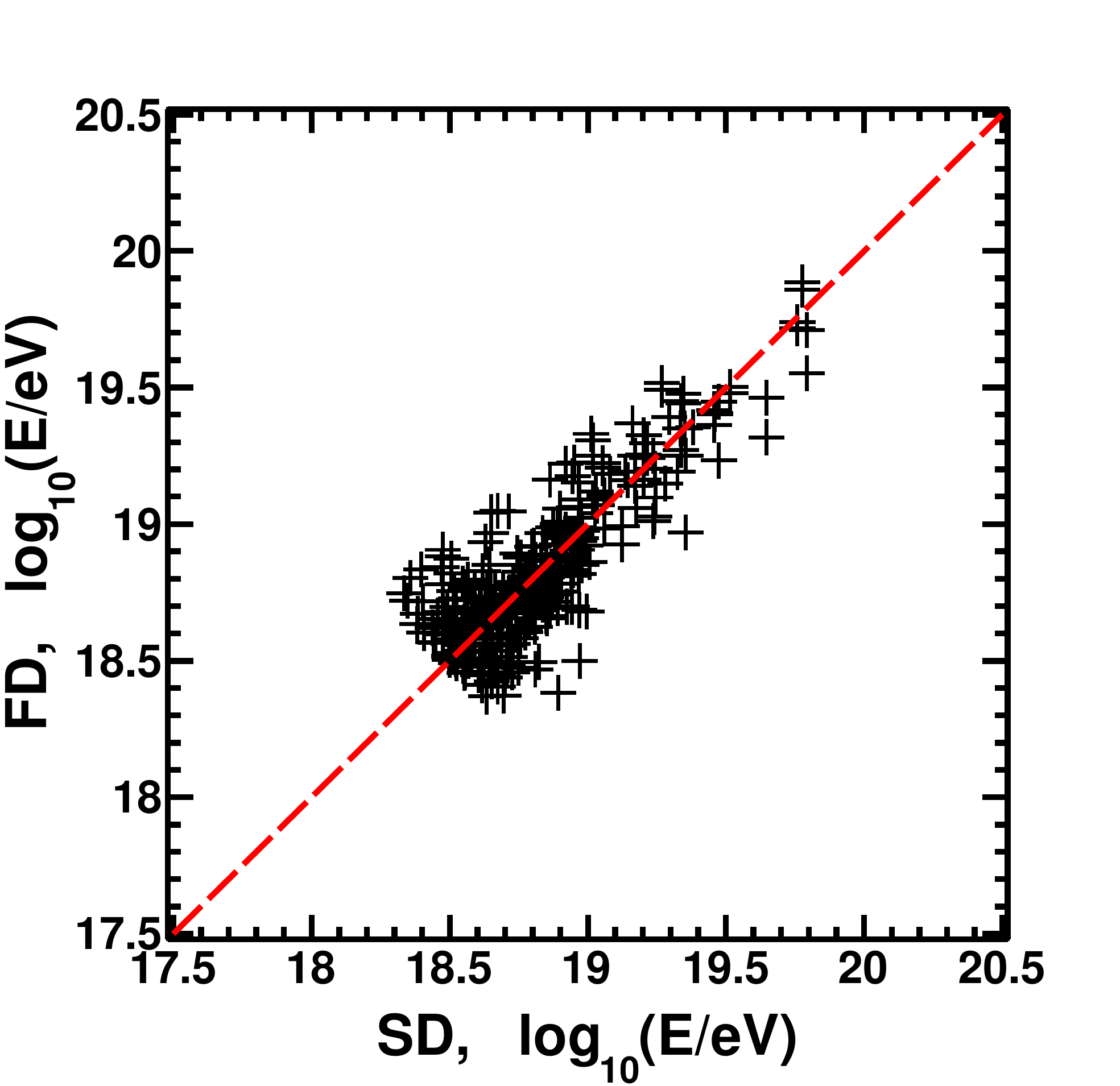}  \includegraphics[width=0.48\textwidth]{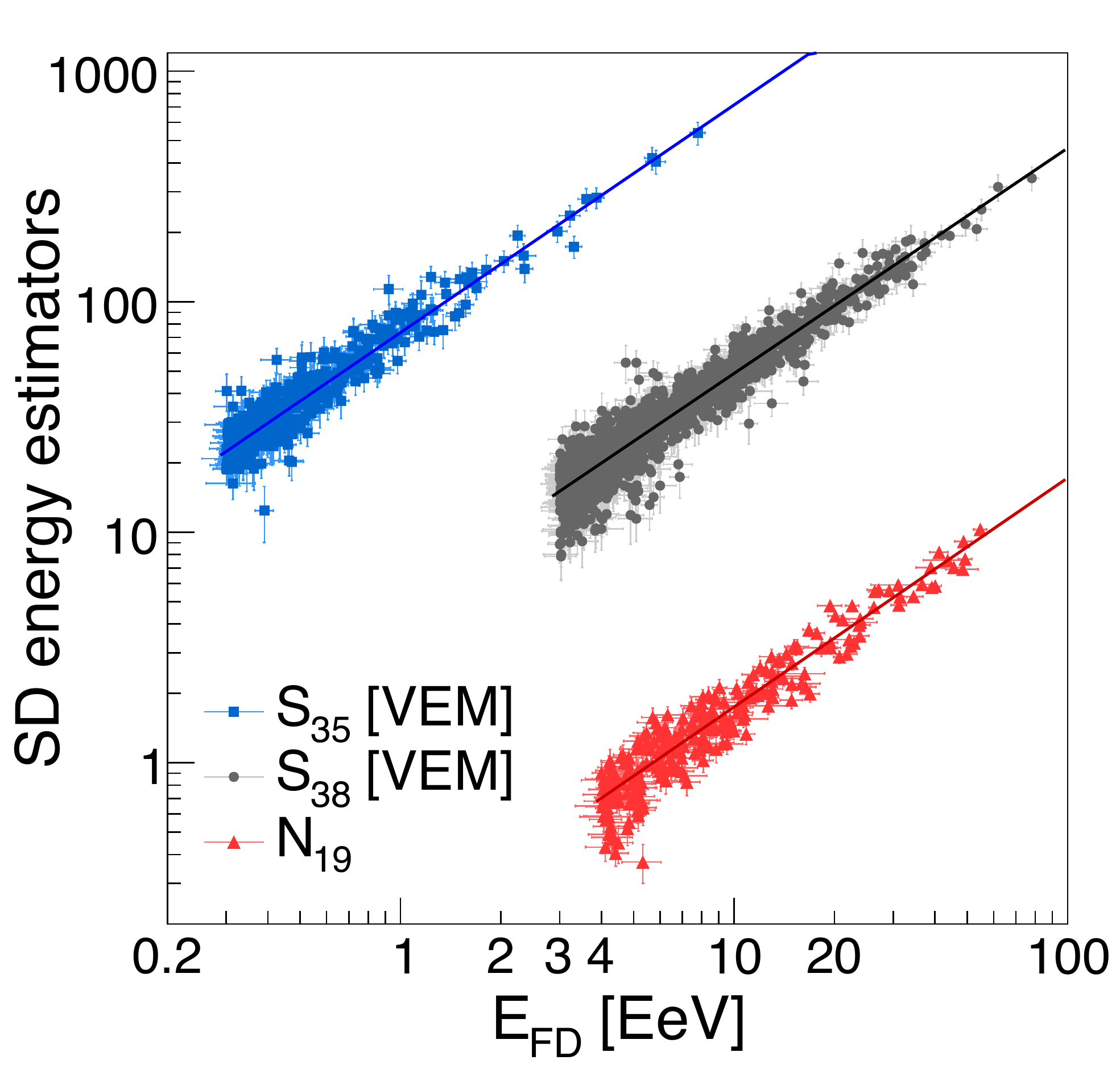}   }
\vspace{0cm}
\caption{Calibration of the SD energy using FD. Left: TA
  analysis~\cite{TA-EnSp-Comb}. For each hybrid event, $E_{\rm FD}$ is
  plotted versus $E_{\rm SDMC}/1.27$ (shown as cross marks).  1.27 is
  the calibration factor that is required to match the SD energies
  estimated using the SD Monte Carlo technique with the energies that
  have been measured by the FD~\cite{TA-EnSp-Comb}.
Right:
  Auger analysis~\cite{Auger-EnSp}.  SD energy estimators $S_{38}$,
  $S_{35}$, and $N_{19}$ are plotted versus corresponding FD energies,
  for the hybrid events relevant for each type of Auger SD
  analysis. Solid lines represents the fits to the power-law function
  described in the text. }
\label{Fig:FDSD}
\end{figure}

In Auger, the reconstruction technique described above is applied to
the showers of $\theta < 60^\circ$ for the 1500 m array and $\theta <
55^\circ$ for the 750 m array.  A different reconstruction technique
is used for the Auger 1500 m array in the case of inclined showers
($\theta > 60^\circ$)~\cite{Auger-HASrec}. In these showers, the
electromagnetic component is largely absorbed by the atmosphere and
the signal in the WCDs is dominated by muons. The muon patterns (maps)
are asymmetric because of the deflections of the
muons in the magnetic field of the Earth. These maps are calculated
for different zenith and azimuth angles using Monte Carlo
simulations. The normalization of the maps, called $N_{19}$, is fitted
to the data and provides an energy estimator for the inclined showers.

\begin{table}[h]
\caption {Quantities Relevant for Calibrating Auger SD Energy Scale~\cite{Auger-EnSp}.}
\label{Table:SDCalibAuger}
\begin{center}
\begin{tabular}{|c|ccc|}\hline 
               & \multicolumn{3}{c|}{Auger SD Analysis Type} \\ 
               & 1500 m Vertical & 1500 m Inclined & 750 m   \\ \hline
Number of Hybrid Events & 1731 & 255 & 469 \\ \hline
$A$ [EeV]     & $0.1871 \pm 0.004$ & $5.71\pm 0.09$ & $(1.29\pm 0.06 )\times 10^{-2}$ \\
$B$               & $1.023\pm 0.006$ & $1.01\pm 0.02$ & $1.01\pm 0.01$  \\ \hline
Energy Resolution & 15.3\%, E $>$ 3 EeV & 19\%, E $>$ 4 EeV & 13\%, E $>$ 0.3 EeV  \\ \hline 
\end{tabular}
\end{center}
\end{table}

\begin{table}[h]
\caption {Quantities Relevant for Calibrating TA SD Energy Scale~\cite{TA-EnSp-Comb,DmitriThesis}.}
\label{Table:SDCalibTA}
\begin{center}
\begin{tabular}{|c|c|}\hline 
Number of Hybrid Events & 551    \\ \hline
$E_{\rm SDMC}/E_{\rm FD}$   & 1/1.27 \\ \hline
                        & \multicolumn{1}{|l|}{19\%,  $E > 10^{19.0}\mathrm{eV}$} \\
Energy Resolution       & \multicolumn{1}{|l|}{29\%,  $10^{18.5}\mathrm{eV} < E < 10^{19.0}\mathrm{eV}$} \\
                        & \multicolumn{1}{|l|}{36\%,  $10^{18.0}\mathrm{eV} < E < 10^{18.5}\mathrm{eV}$} \\ \hline
\end{tabular}
\end{center}
\end{table}

Correlations between the SD energy estimators and the FD energies are
shown in Fig.~\ref{Fig:FDSD}.  In TA, the final event energy is
determined by scaling the result from the Monte Carlo simulations
$E_{\rm SDMC}$ by a factor $\left<E_{\rm SDMC}/E_{\rm FD}\right> =
1.27$, to match the FD energy scale~\cite{TA-EnSp-Comb}.  The same
factor is used for all energies in TA. In Auger, the correlations
between the SD energy estimators and the FD energies are well
described by a power law function $E_\text{FD} = A S^B$, where $S$ is
$S_{38},~ S_{35}, ~{\rm or}~ N_{19}$, depending on the type of the
Auger SD analysis. The parameters $A$ and $B$ are obtained from the
fits to the data~\cite{Auger-EnSp}.  Quantities relevant for the TA
and Auger SD energy calibration are summarized in
Tables~\ref{Table:SDCalibAuger}~and~\ref{Table:SDCalibTA}. The TA and
Auger SD energy resolution is summarized in the last rows of the
tables.

\subsection{Systematic Uncertainties of the Energy Scale}
\label{SubSec:SystUnc}

Since both experiments calibrate their surface detectors to the FDs,
the systematic uncertainties of their energy scales reduce to those of
the FDs.  Therefore, an effort has been made in both TA and Auger
collaborations to understand the uncertainties that affect the
reconstruction of fluorescence detector
events~\cite{TA-EnergyScale2011,TA-SDsp2013,Auger-EnSc}.
Table~\ref{Table:EnScale} shows a summary of the TA and Auger FD
systematic uncertainties in terms of five major contributions:
fluorescence yield, atmospheric modeling, FD calibration,
determination of the longitudinal profile of the shower, and the {\it
  invisible} energy correction.
\begin{table}[h]
\caption {Systematic uncertainties on the energy scale for
  TA~\cite{TA-EnergyScale2011} and Auger~\cite{Auger-EnSc}. For Auger,
  the variation of the uncertainties refers to the energy range
  between $3\times10^{18}$ eV and $10^{20}$ eV.}
\label{Table:EnScale}
\begin{center}
\begin{tabular}{|c|c|c|}\hline 
 \multicolumn{3}{|c|}{Systematic Uncertainties on the Energy Scale} \\ \hline
& TA & Auger \\ \hline
Fluorescence Yield & 11\%  & 3.6\% \\ \hline
Atmosphere & 11\% & 3.4\%$\div$6.2\% \\ \hline
FD Calibration & 10\% & 9.9\% \\ \hline
FD Reconstruction & 9\% & 6.5\% $\div$5.6\% \\ \hline
Invisible Energy  & 5\% & 3\%$\div$1.5\%  \\ \hline
Other Contributions & & 5\% \\ \hline
Total & 21\% & 14\% \\ \hline
\end{tabular}
\end{center}
\end{table}

The fluorescence yield model used by the TA collaboration is based on
a combination of the measurements of the absolute yield by Kakimoto
{\it et al.}~\cite{FY-Kakimoto}, in the 300 to 400 nm range, and the
fluorescence spectrum measured by FLASH~\cite{FY-FLASH}.  Temperature
and pressure dependencies of the absolute FY in TA are taken into account
by the Kakimoto {\it et al.}~\cite{FY-Kakimoto} model also.  Auger uses
all FY measurements performed by the Airfly experiment. Airfly results
include a precise measurement of the absolute intensity at the 337 nm
emission band~\cite{FY-Airfly_AbsYield}, with an uncertainty of $4\%$,
the wavelength spectrum ~\cite{FY-Airfly_spectrum}, and the dependence
on pressure~\cite{FY-Airfly_spectrum}, temperature, and
humidity~\cite{FY-Airfly_T_h,FY-Airfly_Th_Martina} of the emission
bands at different wavelengths.  Contributions of the FY models to the
systematic uncertainty on the energy scale are 11\% for TA and 3.6\%
for Auger. In both cases, the FY model contributions are dominated by
the systematic uncertainties on the absolute FY.

In TA, the aerosol transmission is estimated using a median value of
the aerosol optical depth profiles measured by the
LIDAR~\cite{TALIDAR}.  The uncertainty on the shower energy
determination, obtained by propagating the standard deviation of the
LIDAR measurements, is under 10\%. The Auger collaboration uses hourly
estimates of the aerosol profile provided by the laser facilities
placed in the middle of the SD
array~\cite{Auger-CLF1,Auger-CLF2}. Uncertainties of these
measurements contribute less than 6\% to the reconstructed shower
energy.  A minor contribution to the systematic uncertainty arises
from an imprecise knowledge of the atmospheric density profiles.  Both
TA and Auger use the Global Data Assimilation System (GDAS) that
provides atmospheric data in $1^{\circ} \times 1^{\circ}$ grid points
in longitude and latitude ($\sim 110 ~{\rm km} \times 110 ~{\rm km}$)
all over the world, with a time resolution of 3 h~\cite{GDAS}. A
detailed discussion on the implementation of the GDAS atmospheric
profiles in the Auger FD event reconstruction can be found
in~\cite{Auger-GDAS}.

The uncertainties due to the calibration of the FD telescopes
contribute $\sim$10\% for both TA and Auger. They are dominated by the
uncertainties on the absolute calibration described in
Sec.~\ref{Sec:TAAuger}. The uncertainties from the relative
calibration systems, which allow one to track the short and long term
changes of the detector response, are taken into account in both
experiments, and are small in comparison to those due to the absolute
calibration.

The uncertainties arising from the reconstruction of the longitudinal
shower profiles are obtained by comparing different reconstruction
techniques, as well as from studying the energy reconstruction biases
with Monte Carlo simulations.  Contributions due to the shower profile
reconstruction are $\sim$9\% for TA and $\sim$6\% for Auger.

The last important contribution is the uncertainty due to the
determination of the {\it invisible} energy $E_{\rm inv}$. The TA
collaboration mainly estimates $E_{\rm inv}$ from Monte Carlo
simulations of the proton air showers, using the QGSJetII-03 hadronic
interaction model. For TA, the contribution to the systematic
uncertainty on $E$ due to the missing energy correction is estimated
to be 5\%.  The Auger collaboration derives the invisible energy
correction using data~\cite{Auger-Einv}. This is done by exploiting
the WCDs sensitivity to the muons of the showers.  The muons are
mostly originating from the pion decays, with an associated muon
neutrino (or muon antineutrino), and therefore, the signal in the WCDs
is sensitive to the muon size of the shower, and it is well correlated
with the $E_{\rm inv}$. This analysis allows to keep the uncertainty
from the {\it invisible} energy estimate on the Auger energy scale
well under 3\%.  Recently, the TA collaboration has also performed a
check of the missing energy calculation by using inclined showers of
the data, following this method ~\cite{Auger-Einv}.

The total uncertainty on the energy scale is obtained by adding in
quadrature all individual contributions.  It is found to be 21\% for
TA, and 14\% for Auger.  In addition, for Auger, the total uncertainty
includes a further contribution of 5\%, which has been evaluated by
studying the stability of the energy scale in different time periods
and/or under different conditions.

\subsection{Energy Scale Comparison between the TA and Auger}
\label{Sec:EnergyScaleComparison}

Understanding all contributions to the difference in the energy scale
between the two experiments is a difficult task, since many factors
are related to the performance of the detectors and to the differences
of the analyses techniques used by the two collaborations. On the
other hand, two important contributions, the fluorescence yield and
the invisible energy, can be considered as external parameters of the
experiments, in the sense that they are related to the general
properties of the atmospheric showers and thus they can be easily
implemented in the CR event reconstruction chains in both
collaborations. Therefore, the difference in the energy assignment can
be addressed by studying the differences in the fluorescence yield
(FY) model and in the {\it invisible} energy corrections.

\begin{figure}[h]
\centerline{
\includegraphics[width=0.47\textwidth]{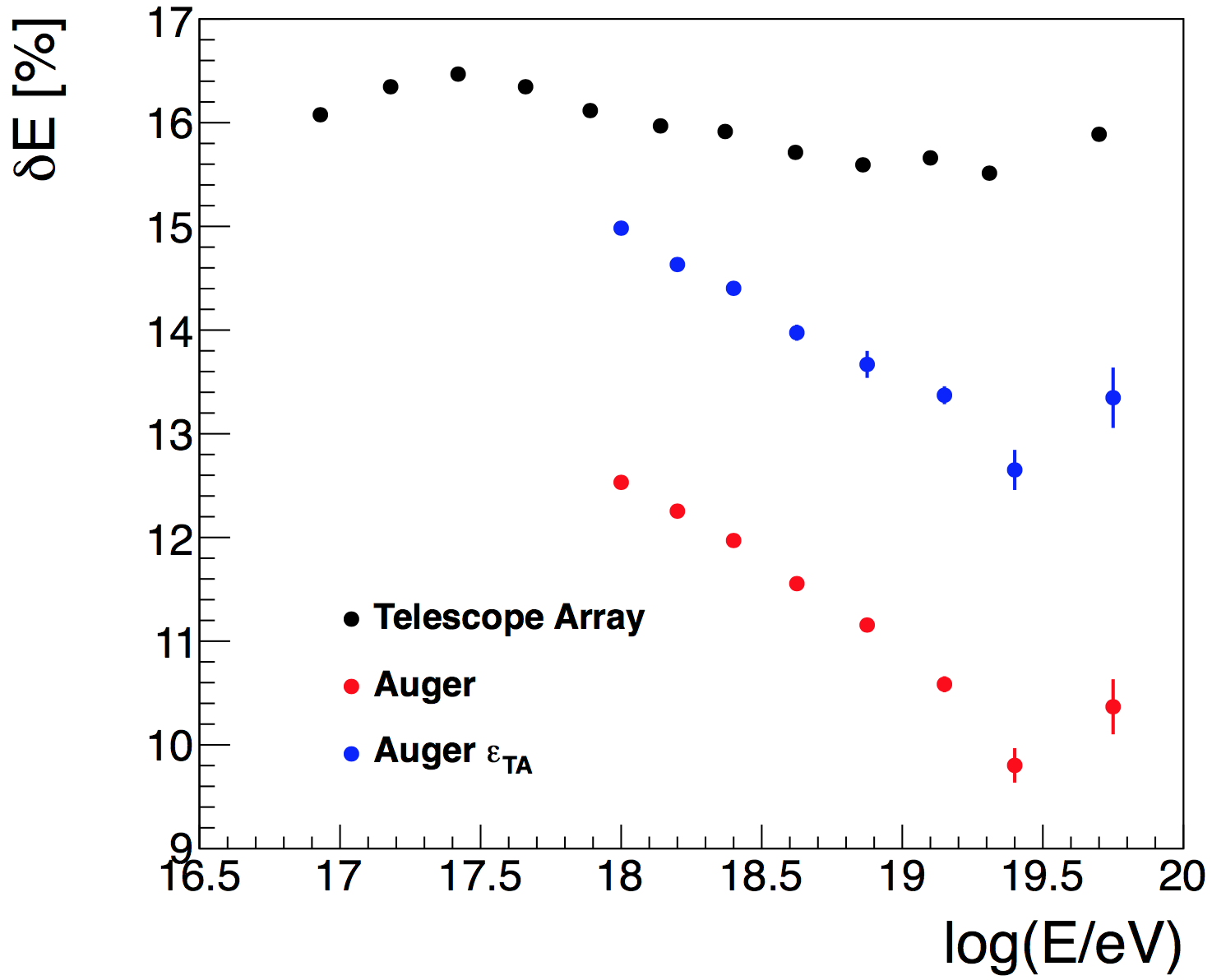}
\includegraphics[width=0.53\textwidth]{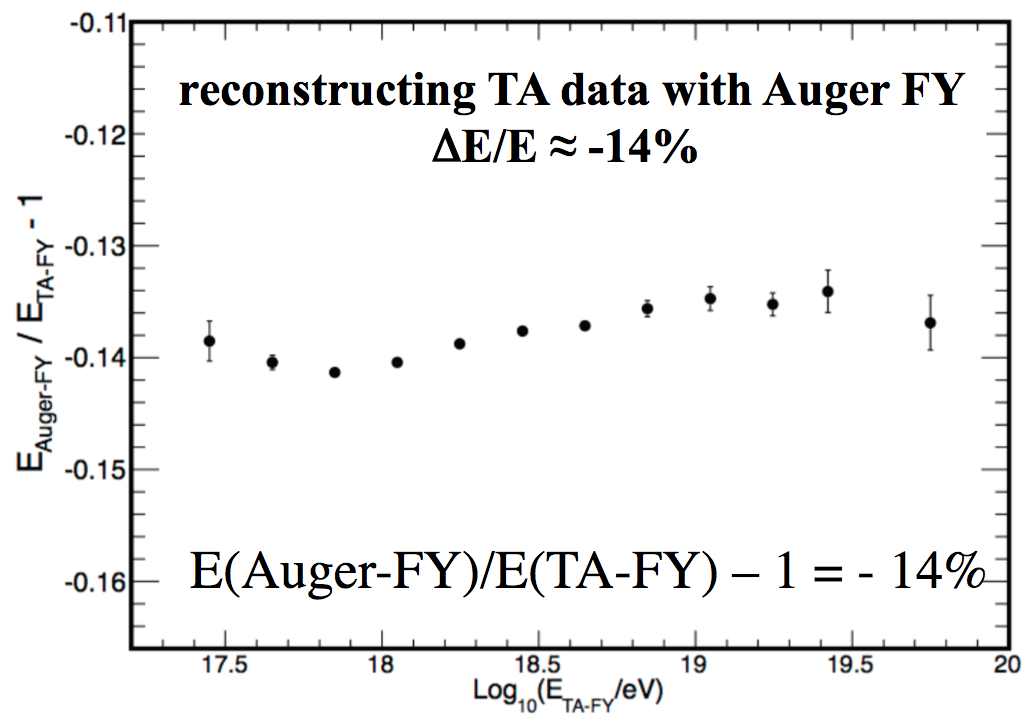}
}
\caption{Effect of the change of the fluorescence yield (FY) in the reconstruction of the FD events. 
Right: shift of TA energies when the Auger FY is used. 
Left: the shift of the Auger energies when the TA FY is implemented
is shown with red points and the blue points refer to when 
the effect of the different spectral responses of Auger and TA telescopes is
taken into account~\cite{JoseRamonThesis}. The inverse of the TA energy shift 
of the right figure (E(TA-FY)/E(Auger-FY)-1) is shown with black points.}
\label{Fig:EnergyShifts-FY}
\end{figure}

The impact of the FY on the reconstruction of the fluorescence events
has been studied in detail since many
years~\cite{WG1,JoseRamon-icrc13,JoseRamonThesis}. In the left panel
of Figure~\ref{Fig:EnergyShifts-FY}, we report the results of the
studies performed in~\cite{JoseRamonThesis}. Red points describe the
effect (on Auger shower energies) of changing the fluorescence yield
model from the FY model used by Auger to the FY model used by TA. The
energy shift is $\sim12\%$ at 1 EeV and is slightly smaller at the
highest energies.

This energy shift is the result of the combined effect to change the
absolute intensity of the fluorescence yield and all parameters
describing the relative intensities of the spectral lines and their
dependence on the atmospheric conditions. The effects of the single
components can be disentangled by the following argument. The absolute
FY from Kakimoto et al. ~\cite{FY-Kakimoto}, when normalized to the
intensity of the 337 nm line, where the Airfly experiment made a
precise measurement of the absolute FY, differs from that of the
Airfly measurement~\cite{FY-Airfly_AbsYield} by $\sim20\%$ (Airfly FY
is higher)~\cite{ArquerosLast}. If the absolute FY from Kakimoto et
al.~\cite{FY-Kakimoto} was used in the reconstruction of the Auger
events, while retaining all other parameters of the Airfly model, one
would expect the Auger energies to increase by $\sim20\%$. From this,
we conclude that the effects of the FY parameters, other than the
absolute FY, are of the order of $- 10\%$. About half of this effect
is due to the removal of the temperature and humidity dependence of
the quenching cross sections (see also~\cite{Auger-atm-showers}),
effects that are properly accounted for in Auger experiment. We note
that the $20\%$ difference between the Kakimoto et al. and Airfly
absolute FYs is outside of the range defined by the uncertainties
stated by the two measurements, $10\%$~\cite{FY-Kakimoto} and
$3.9\%$~\cite{FY-Airfly_AbsYield}, respectively.

The right panel of the Figure~\ref{Fig:EnergyShifts-FY} describes the
effect of changing the fluorescence yield model in the reconstruction
of the fluorescence detector events seen by TA~\cite{WG2}. If TA were
to use the FY model of Auger, the TA energy scale would be reduced by
$\sim 14\%$. The inverse of this energy shift is directly comparable
with the energy shift that is expected in the case of Auger using the
TA FY, as shown in the left panel of the Figure~\ref{Fig:EnergyShifts-FY}
using black points.

It is not surprising that the $\Delta E/E$ results of the TA and Auger
(black and red points in figure~\ref{Fig:EnergyShifts-FY} on the left)
are different. For each experiment, the spectrum of the fluorescence
photons detected by the FD is necessarily different from the one
emitted at the axis of the cosmic ray shower: the fluorescence photon
spectrum is folded with the FD spectral response, and the atmospheric
transmission also dependents on the wavelength. Since the Auger and TA
FD spectral responses and atmospheric transmission conditions are
generally different, we expect larger differences for the higher
energy showers that are occurring farther away from the telescopes. A
better agreement between the energy shifts can be obtained by
correcting the Auger energy shift for the effects due to the different
spectral response. The results of this analysis are shown in the left
panel of Figure~\ref{Fig:EnergyShifts-FY}~\cite{JoseRamonThesis} with
blue dots, which are now in a better agreement with the TA energy shift
(black points).

Following the above studies we conclude that, despite the above
mentioned inconsistency between the Airfly~\cite{FY-Airfly_AbsYield}
and Kakimoto et al.~\cite{FY-Kakimoto} absolute FYs, the difference in
the energy scales of TA and Auger due to the use of a different FY
model are at the level of $10 -15\%$ and are roughly consistent with
the estimated uncertainties presented in Sec.~\ref{SubSec:SystUnc}.

The validity of the estimations of the uncertainties on the FY has
been also addressed by the ELS facility at the TA experiment.
Preliminary results of several ELS runs, under different atmospheric
conditions, have been presented in~\cite{ELS-ICRC15}.  The ELS results
are in a better agreement with the Airfly FY model.

\begin{figure}[h]
\centering
\includegraphics[width=0.6\textwidth]{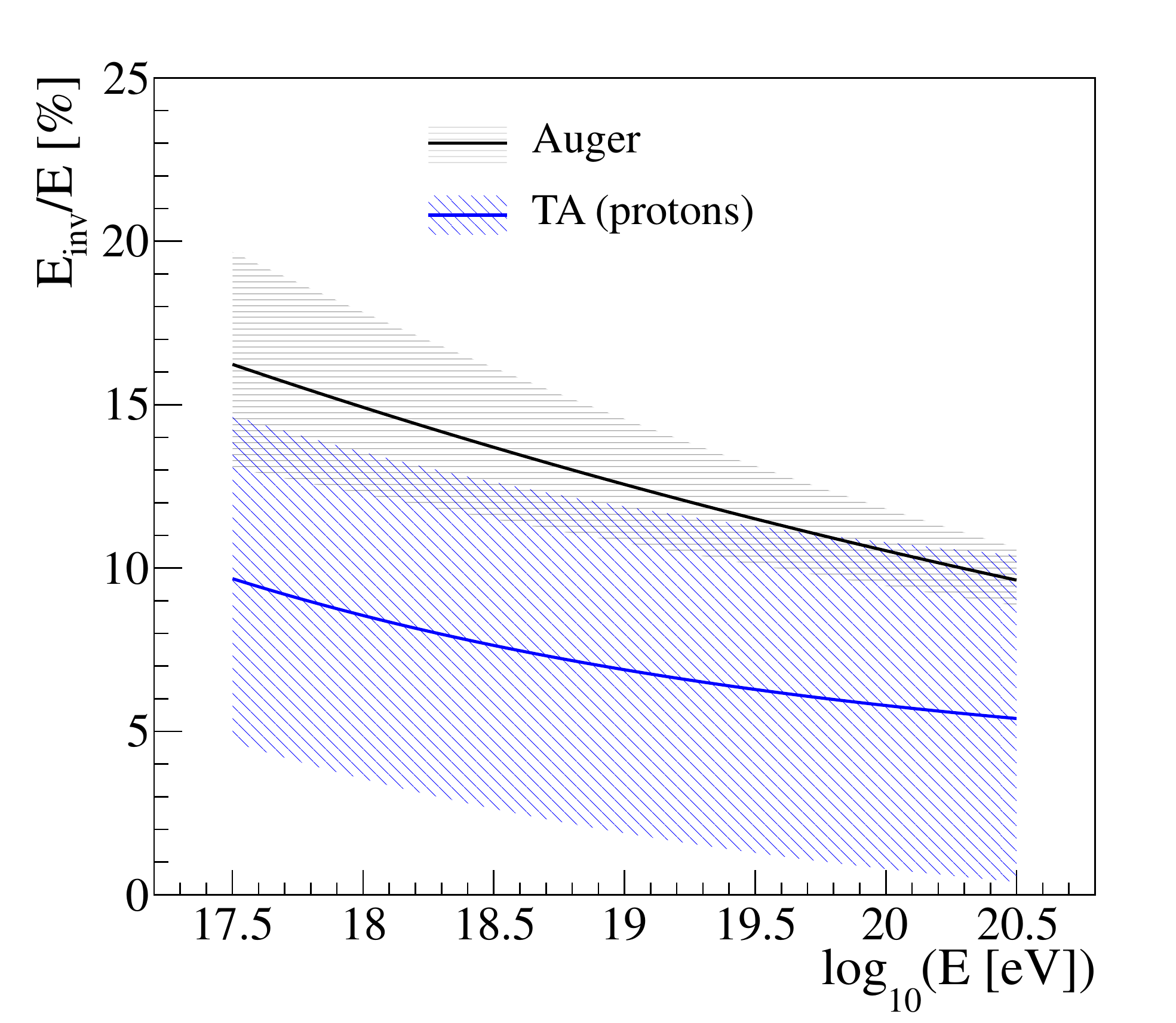}
\caption{Invisible energy contribution to the total shower energy in the reconstruction of Auger and TA fluorescence events~\cite{TA-EnergyScale2011,TA-SDsp2013,Auger-EnSc}.}
\label{Fig:Einv}
\end{figure}

The {\it invisible} energy (E$_{inv}$) corrections implemented by the
TA and Auger are shown in
Figure~\ref{Fig:Einv}~\cite{TA-EnergyScale2011,TA-SDsp2013,Auger-EnSc}. They
are presented in terms of the percent contribution to the total shower
energy $E$.  At $10^{19}$ eV, the TA invisible energy correction is
7\%, while that of the Auger is 13\%.  The difference between the two
corrections is about 6\% (slightly smaller at higher energies).  The
two corrections agree within the systematic uncertainties quoted by
the two collaborations that are shown using dashed bands in
Fig.~\ref{Fig:Einv}.

As already addressed in Sec.~\ref{SubSec:SystUnc} the {\it invisible}
energy of TA has been estimated using Monte Carlo simulations of
proton primaries with the QGSJetII-03 hadronic interaction model.
For a heavier composition, the invisible energy correction would be
larger.  The assumption of the proton primaries is consistent with the
light composition observed by TA through the measurement of the mean
value of the maximum of the shower development ($\langle
X_{\mathrm{max}} \rangle$)~\cite{TA-mass}. It is worthwhile noting
that the inference on mass composition strongly depends on the
hadronic interaction models used to interpret $\langle
X_{\mathrm{max}} \rangle$~\cite{Unger-rev}.  The Auger measurements of
$\langle X_{\mathrm{max}} \rangle$~\cite{Auger-mass} are consistent
with those of the TA~\cite{TA-Auger-mass}, but they generally support
a heavier mass composition.


The Auger {\it invisible} energy correction has the advantage to be
essentially insensitive to the hadronic interaction models since it is
derived from the data.  It has rather high values, even higher than
the one predicted by the simulations for iron primaries.  These higher
values are due to the excess of muons measured by Auger in highly
inclined events~\cite{Auger-muons}. 

We can conclude this section by estimating the energy shifts of the
Auger and TA energy scales by changing both the FY and the {\it
  invisible} energy. As a first approximation they can be obtained by
combining the two energy shifts previously presented.  The energies of the TA
events would be decreased
by about 9\% ($(1-14\%) \times (1+6\%) = -0.91$) while the energies of
Auger would be increased by about +5\% ($(1+12\%) \times (1-6\%) = 1.05$).

\section{TA and Auger Energy Spectrum}
\label{Sec:Spectrum}

Energy spectrum is obtained by dividing the energy distribution of
cosmic rays by the accumulated exposure of the detector.  The
calculation of the exposure for the surface detector is generally
robust, especially above the energy threshold where the array is fully
efficient regardless of the event arrival direction.  For the
fluorescence detector, on the other hand, the calculation of the
exposure should take into account the detector response as a function
of energy and distance between the shower and the telescope,
conditions of the data collection, and the state of the atmosphere.
Large exposures accumulated by the surface detectors of Auger and TA
experiments make it possible to study the UHECR flux at very high
energies in different declination bands, and the measurements can be
used to constrain the astrophysical models. 

\subsection{Energy Spectrum Measurement}
\label{Sec:SpectrumMeasurement}

\subsubsection{TA Data}
\label{subsec:TAdata}

\begin{figure}[h]
\centerline{\includegraphics[width=1.0\textwidth]{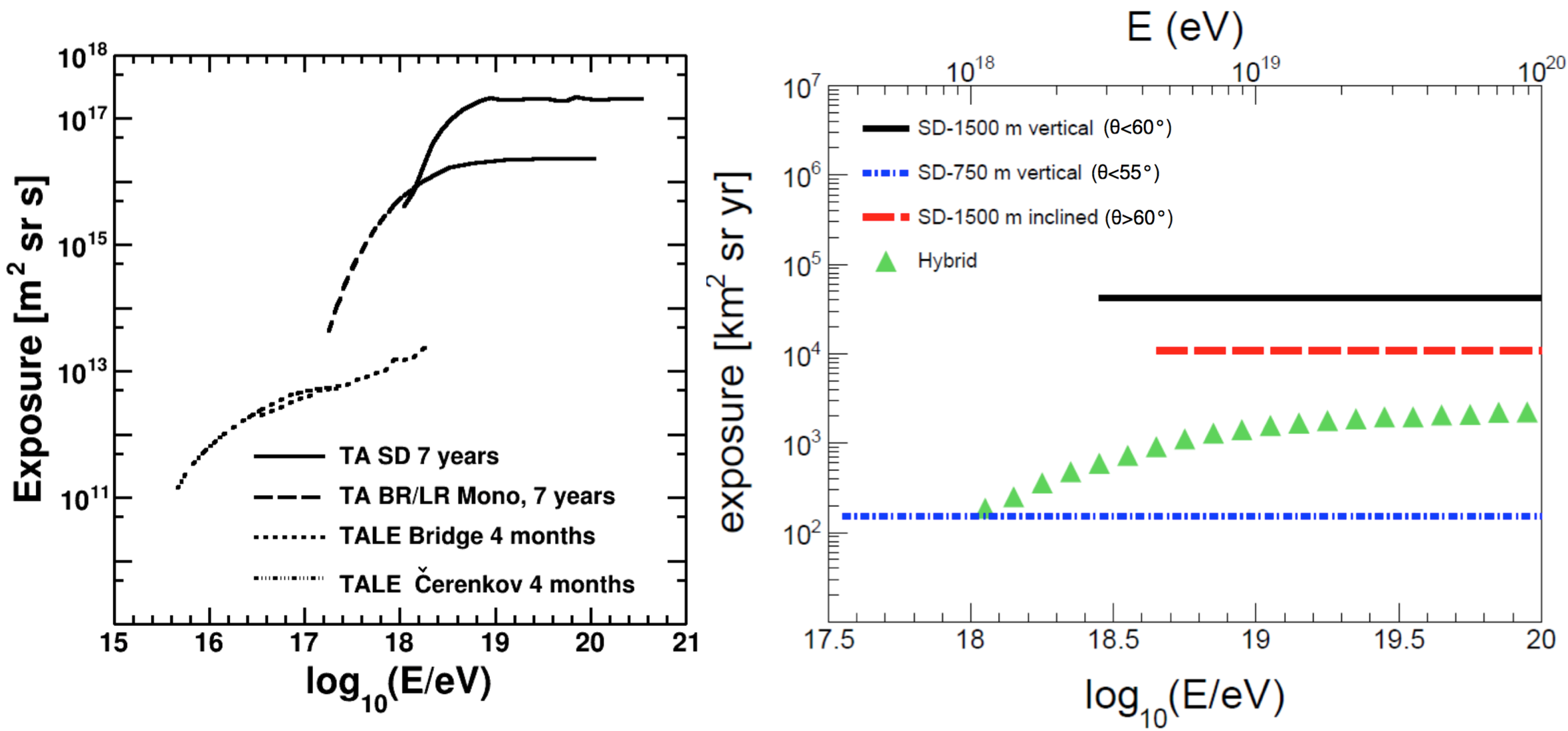}}
\caption{Exposure for the energy spectrum measurements of TA (left
  panel)~\cite{TA-EnSp-Comb} and Auger (right
  panel)~\cite{Auger-EnSp}.  The TA exposure is stated in units of
  $[$m$^2$ sr s$]$ while the Auger exposure is shown in units of
  $[$km$^2$ sr yr$]$. The multiplication factor to obtain the TA
  exposure in $[$km$^2$ sr yr$]$ units is $3.2\times 10^{-14}$.}
\label{Fig:Exposure_TA_Auger}
\end{figure}

\begin{figure}[h]
\centerline{\includegraphics[width=0.95\textwidth]{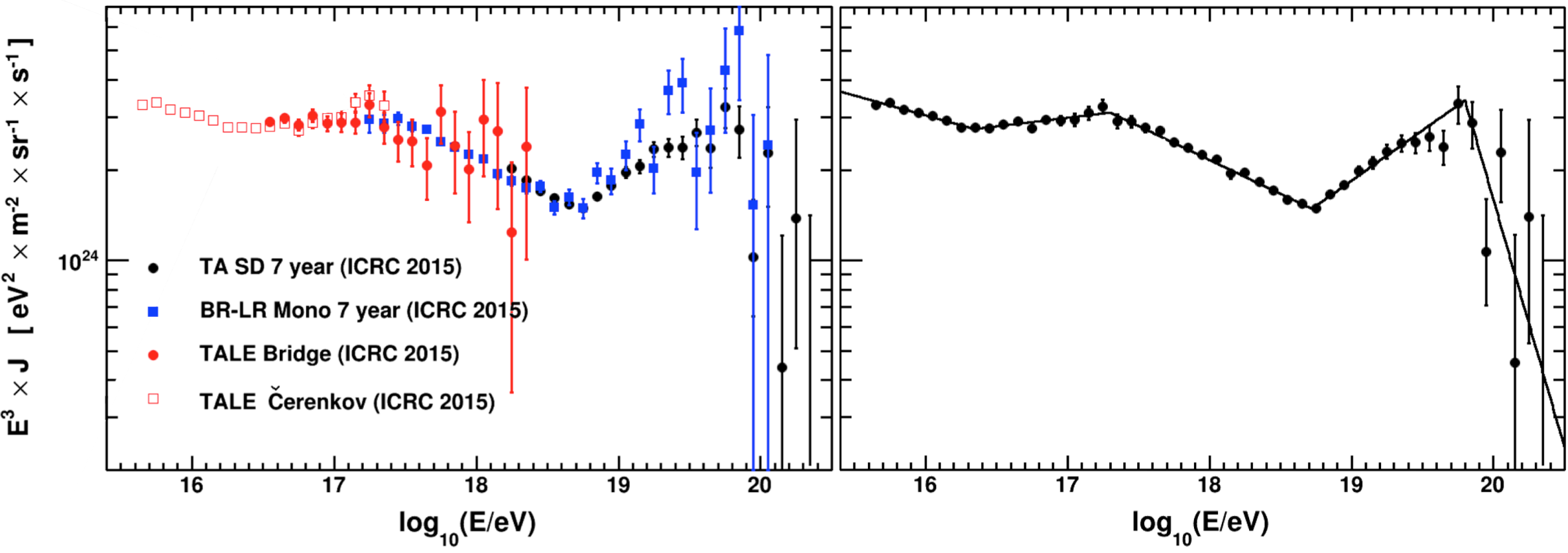}}
\caption{Energy spectra presented by the TA collaboration at the 34rd International Cosmic Ray Conference (ICRC 2015)~\cite{TA-EnSp-Comb}.}
\label{Fig:EnSp_TA}
\end{figure}

~\\
The TA collaboration has measured four independent energy
spectra~\cite{TA-EnSp-Comb}. The highest energies are covered by the
SD, the intermediate energies are covered by the BR and LR
telescopes~\cite{TA-EnSp-Mono,TA-EnSp-Mono-paper}, and the lowest
energies are measured by the TALE telescopes using Cherenkov light.
The TALE events have been divided into two categories, one in which
the fluorescence light is dominating the flux of photons detected by
the telescopes (TALE Bridge)~\cite{TA-icrc15-TALE2}, and another one
where the Cherenkov light is the dominant component (TALE
Cherenkov)~\cite{TA-icrc15-TALE1}.  The exposures for the four
different reconstruction methods are shown in the left panel of
Fig.~\ref{Fig:Exposure_TA_Auger} and the energy spectra are shown in
the left panel of Fig.~\ref{Fig:EnSp_TA}. The spectrum obtained by
combining the four measurements is presented in the right panel of
Fig.~\ref{Fig:EnSp_TA}.    

The TA spectrum, including the TA low energy extension, covers over
4.7 orders of magnitude in energy, starting at $4 \times 10^{15}$ eV,
just above the {\it knee}. The analysis of the TALE data has allowed
to observe the low energy {\it ankle} at $\sim 2 \times 10^{16}$ eV
and the second {\it knee} at $\sim 2 \times 10^{17}$ eV. The {\it
  ankle} and the cut-off of the UHECR spectrum are confirmed with the
improved statistics by the BR and LR FD and by the SD.  At the very
high energies the combined spectrum is dominated by the SD
measurements. The TA SD is fully efficient above $8 \times 10^{18}$ eV
and its energy scale is fixed by the FD as described in
Sec.~\ref{SubSec:SDCalib}.  The TA SD exposure accumulated over 7
years
of data taking is $\sim$6300 km$^2$ sr yr. This is estimated using a
detailed Monte Carlo simulation that takes into account the detector
effects and includes the unfolding corrections that have to be applied
to the observed event energy distribution to take into account the
bin-to-bin migrations due to the finite resolution of the
detector~\cite{DmitriThesis}.  Due to the steepness of the spectrum,
the effects of the resolution would otherwise be causing a positive
bias in the observed flux, since the upward fluctuations of the
energies are not fully compensated by downward fluctuations.

It is customary to characterize the shape of the spectra using
suitable functional forms.  As seen in Fig.~\ref{Fig:EnSp_TA},
the TA collaboration uses power laws with break points that correspond
to the energies at which the spectral indexes change their values.
Above $\sim 3 \times 10^{17}$ eV the function is expressed as 
\begin{eqnarray*}
J(E) & \propto & E^{-\gamma_1}   \hspace{1cm}  E < E_{\rm ankle} \\
J(E) & \propto & E^{-\gamma_2}   \hspace{1cm}  E_{\rm ankle} < E < E_{\rm break}\\
J(E) & \propto & E^{-\gamma_3}   \hspace{1cm}  E > E_{\rm break} 
\end{eqnarray*}
and the values of the fitted parameters are shown in Table~\ref{Table:FitSpectrum_TA}.
\begin{table}[h]
\caption {Values of the broken power law fit parameters for the TA energy spectrum above $\sim 3 \times 10^{17}$ eV~\cite{TA-EnSp-Comb}. Only statistical uncertainties are shown.}
\label{Table:FitSpectrum_TA}
\begin{center}
\begin{tabular}{|c|c|c|c|c|}\hline 
            \multicolumn{5}{|c|}{TA spectrum parameters} \\ \hline
      $\gamma_1$ & $\gamma_2$ & $\gamma_3$ & $E_{\rm ankle}$ [EeV] & $E_{\rm break}$ [EeV]\\ 
      $3.226 \pm 0.007$ & $2.66 \pm 0.02$ & $4.7 \pm 0.6$ &  $5.2 \pm 0.2$ & $60 \pm 7$ \\ \hline
\end{tabular}
\end{center}
\end{table}

\subsubsection{Auger Data}
\label{subsec:Augerdata}

\begin{figure}[h]
\centerline{
\includegraphics[width=0.48\textwidth]{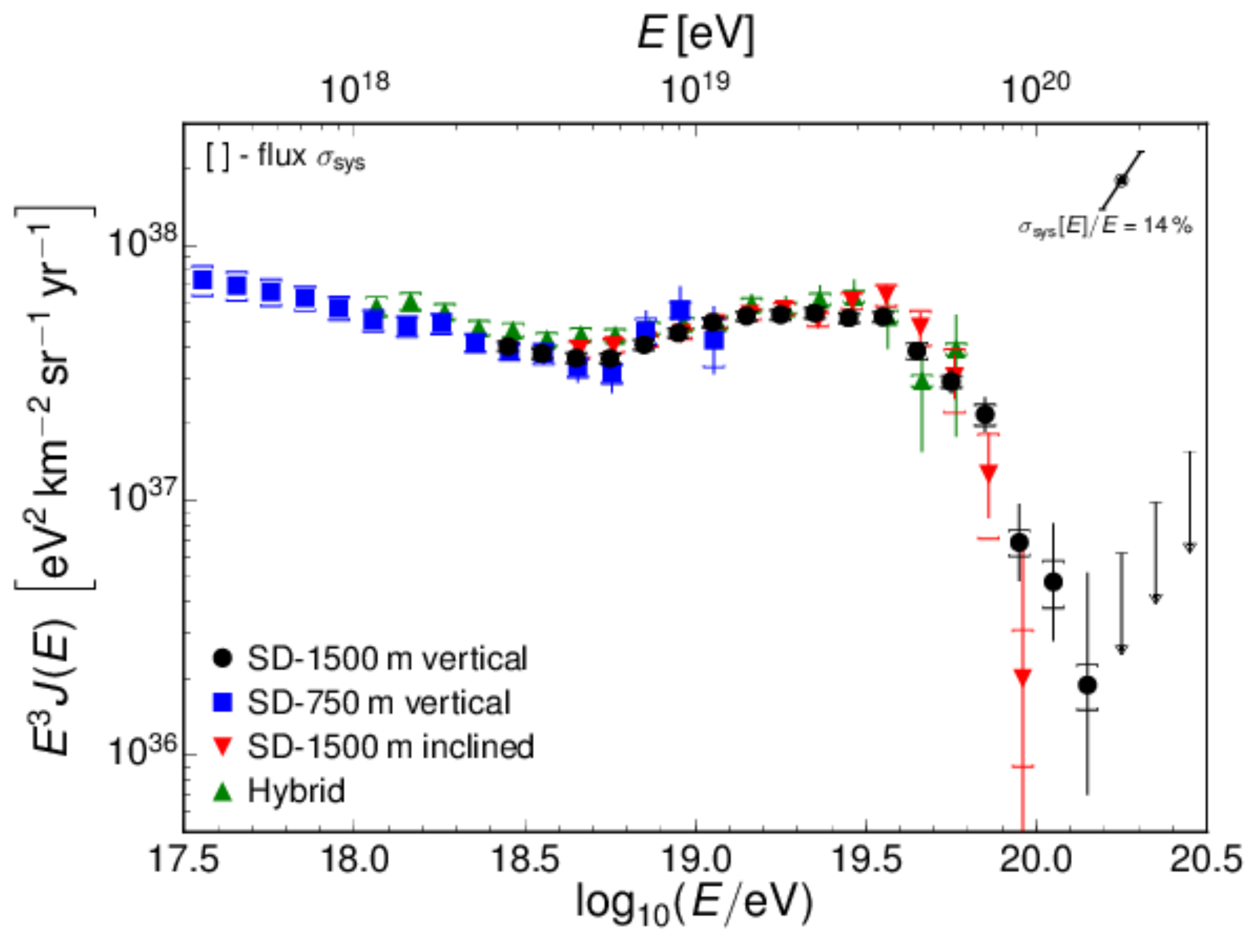} 
\includegraphics[width=0.52\textwidth]{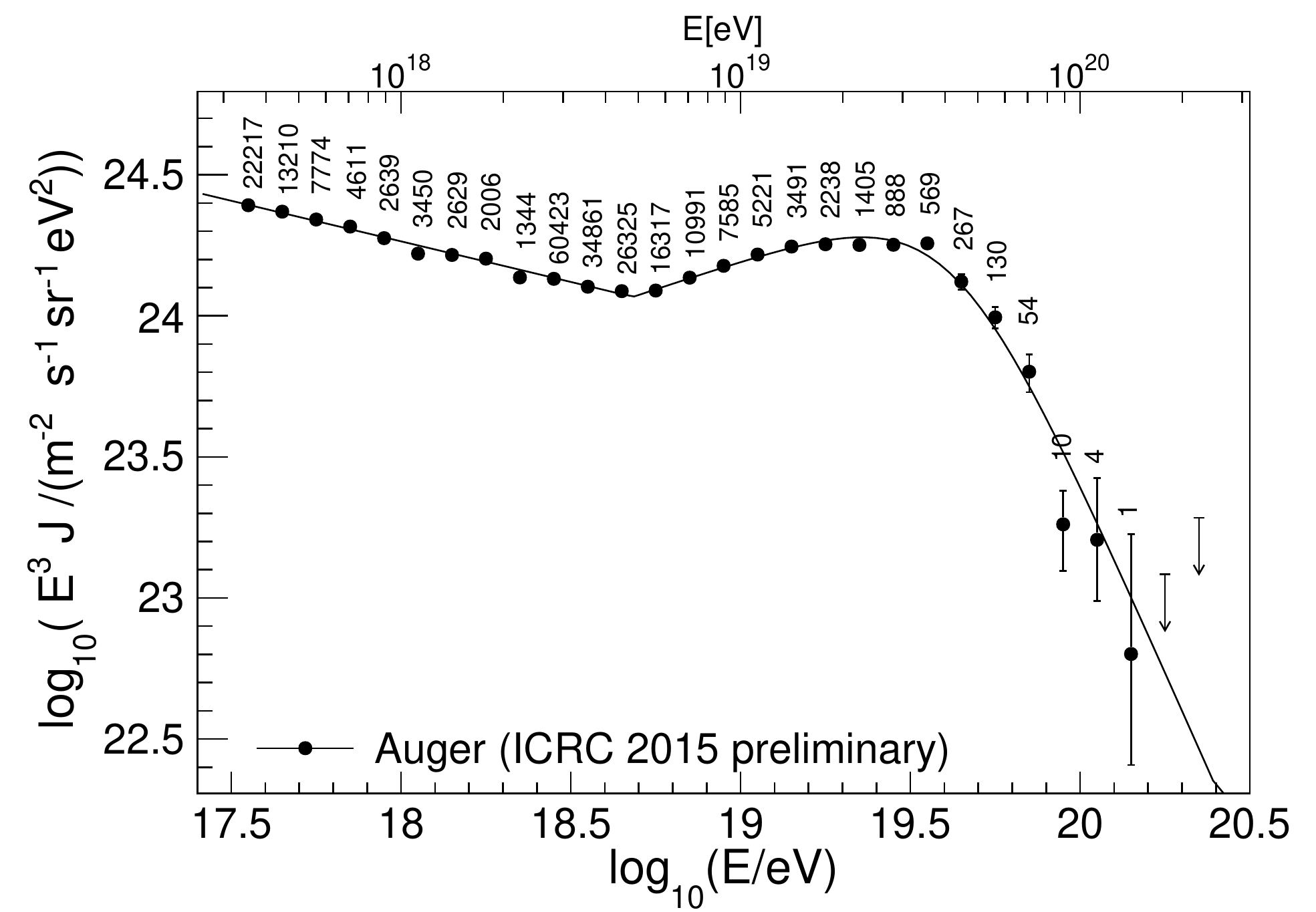}
}
\caption{Energy spectra presented by the Auger collaboration at the 34rd International Cosmic Ray Conference (ICRC 2015)~\cite{Auger-EnSp}. }
\label{Fig:EnSp_Auger}
\end{figure}

~\\
Auger collaboration has measured the energy spectrum using four
different techniques. The first two measurements cover the highest
energies.  The measurements consist of two data sets of vertical and
inclined events seen by Auger surface detector array of 1500 m
spacing.  Intermediate energies are covered by a set of hybrid events
seen by the Auger FD and SD.  Auger 750 m spacing array covers the low
energies down to $3 \times 10^{17}$ eV. The energy calibration of
these showers is done using the fluorescence detector, as explained in
Sec.~\ref{SubSec:SDCalib}.  The four energy spectra are shown in the
left panel of Fig.~\ref{Fig:EnSp_Auger} and the corresponding
exposures are shown in the right panel of
Fig.~\ref{Fig:Exposure_TA_Auger}. The energy spectrum obtained by
combining all four measurements is presented in the right panel of
Fig.~\ref{Fig:EnSp_Auger}.

Large size of Auger water tanks as well as the overall surface area
coverage are the key factors that enabled the Auger collaboration to
perform a high precision measurement of the UHECR energy spectrum with
relatively high statistics.  All four Auger spectra overlap in the
region of the {\it ankle}. The cut-off is precisely measured by the
1500 m array with an exposure of 42500 km$^2$ sr yr for the vertical
and 10900 km$^2$ sr yr for the inclined showers. The data covers a
period of about 10 years. The SD exposure is a purely geometrical
quantity, which is based on the calculation of the number of active
elemental hexagon cells of the array as a function of time, with an
uncertainty of better than 3\%~\cite{Auger-Exposure}.

As can be seen in Fig.\ref{Fig:EnSp_Auger}, Auger collaboration
characterizes the energy spectrum using a functional form that is
different from that used by the TA.  The function used by Auger
consists of a power law below the ankle and a power law with a smooth
suppression at the highest energies
\begin{eqnarray*}
J(E) & = & J_0 \left(  \frac{E}{E_{\rm ankle}} \right)^{-\gamma_1}       \hspace{7.4cm}   E < E_{\rm ankle} \\
J(E) & = & J_0 \left(  \frac{E}{E_{\rm ankle}} \right)^{-\gamma_2}   \left[ 1 +  \left( \frac{E_{\rm ankle}}{E_{s}} \right)^{\Delta\gamma} \right]   \left[ 1 + \left( \frac{E}{E_{s}} \right)^{\Delta\gamma}  \right]^{-1}        \hspace{1cm}  E > E_{\rm ankle} \
\end{eqnarray*}
Here $\gamma_1$ and $\gamma_2$ are the spectral indexes below and
above $E_{\rm ankle}$, respectively, and therefore they have the same
meaning as the corresponding TA parameters.  $E_{s}$ is, with a good
approximation, the energy at which the spectrum drops to a half of
what would be expected in the absence of the cutoff, and
$\Delta\gamma$ is the increment of the spectral index beyond the
suppression region. $ J_0 $ is the overall normalization factor, that
is conventionally chosen to be the value of the flux at $E = E_{\rm
  ankle}$.  The values of the parameters are shown in
Table~\ref{Table:FitSpectrum_Auger}.
\begin{table}[h]
\caption {Values of the parameters of the functional form that fitted to the combined Auger energy spectrum~\cite{Auger-EnSp}. Statistical and systematic uncertainties are shown.}
\label{Table:FitSpectrum_Auger}
\begin{center}
\begin{tabular}{|c|c|c|c|}\hline 
            \multicolumn{4}{|c|}{Auger spectrum parameters} \\ \hline
$J_{0}$ [eV$^{-1}$km$^{-2}$sr$^{-1}$yr$^{-1}$]     & $\gamma_1$                          & $\gamma_2$                        & $\Delta\gamma$ \\ 
$(3.30 \pm 0.15 \pm 0.20) \times 10^{-19}$         & $ 3.29 \pm 0.02 \pm 0.05 $   & $ 2.60 \pm 0.02 \pm 0.1 $  &  $3.1 \pm 0.2 \pm 0.4$  \\ \hline
\multicolumn{2}{|c|}{ $E_{\rm ankle}$ [EeV]}    & \multicolumn{2}{c|}{ E$_{s}$ [EeV]}    \\ 
\multicolumn{2}{|c|}{ $4.82 \pm 0.07 \pm 0.8$}    & \multicolumn{2}{c|}{ $42.1 \pm 1.7 \pm 7.6$ }    \\ \hline
\end{tabular}
\end{center}
\end{table}

Auger SD spectrum is corrected for the effects of the detector
resolution using a forward-folding approach.  First, a Monte-Carlo
simulation of the detector is used to calculate the resolution
bin-to-bin migration matrix.  Next, the measured Auger spectrum
(before it has been corrected for the effects of the resolution) is
fitted to the convolution of the functional form (described above) and
the bin-to-bin migration matrix.  Once the best fit parameters
(Table~\ref{Table:FitSpectrum_Auger}) are obtained, the resolution
correction factor is calculated by dividing the fitted spectrum
function by the convolution of the fitted spectrum function and the
bin-to-bin migration matrix.  The final Auger spectrum result is
obtained by applying this resolution correction factor to the initial
measurement of the spectrum.

\subsection{Comparison of the TA and Auger Results}
\label{Sec:SpectrumComparison}

\begin{figure}[h]
\centerline{
\includegraphics[width=0.75\textwidth]{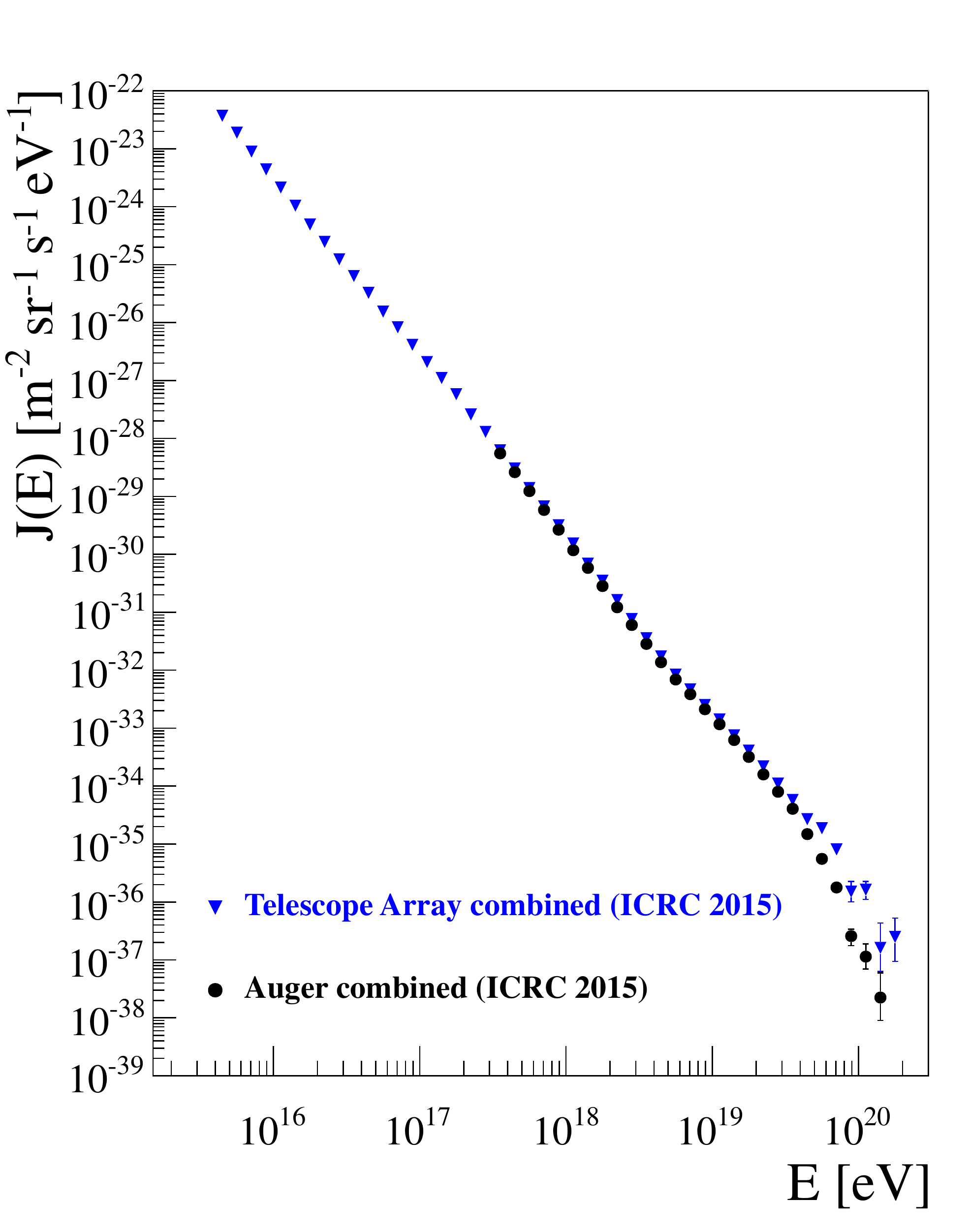} 
}
\caption{The TA and Auger combined energy spectra $J(E)$ as a function
  of $E$ presented at the 34rd International Cosmic Ray Conference
  (ICRC 2015)~\cite{TA-EnSp-Comb,Auger-EnSp}.}
\label{Fig:EnSp_AugerTA_JvsE}
\end{figure}

It is customary, in both TA and Auger, to present the cosmic ray
spectrum as flux $J(E)$ multiplied by the third power of energy
($E^3$) (see Fig.~\ref{Fig:EnSp_TA} and~\ref{Fig:EnSp_Auger}).  In
this representation, the {\it low energy ankle} and the {\it ankle}
are clearly seen as the local minima, while the {\it second knee} and
the high energy suppression appear as the local maxima.
Figure~\ref{Fig:EnSp_AugerTA_JvsE} shows superimposed TA and Auger
spectra simply as $J(E)$ vs $E$.  Stronger features, {\it ankle} and
the suppression, are still seen in the two results, even without
multiplying them by $E^{3}$.

\begin{figure}[h]
\centerline{
\includegraphics[width=0.5\textwidth]{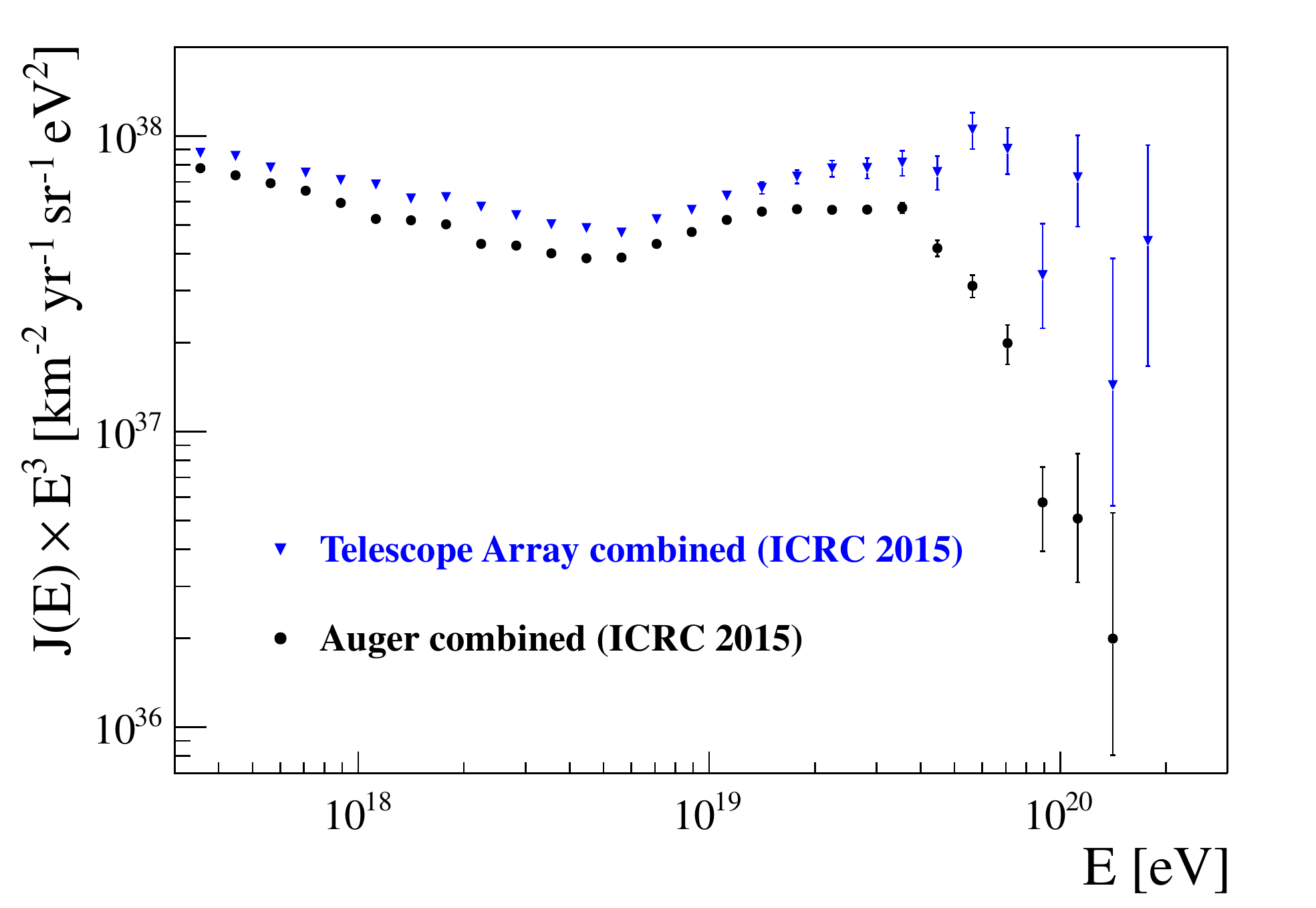} \includegraphics[width=0.5\textwidth]{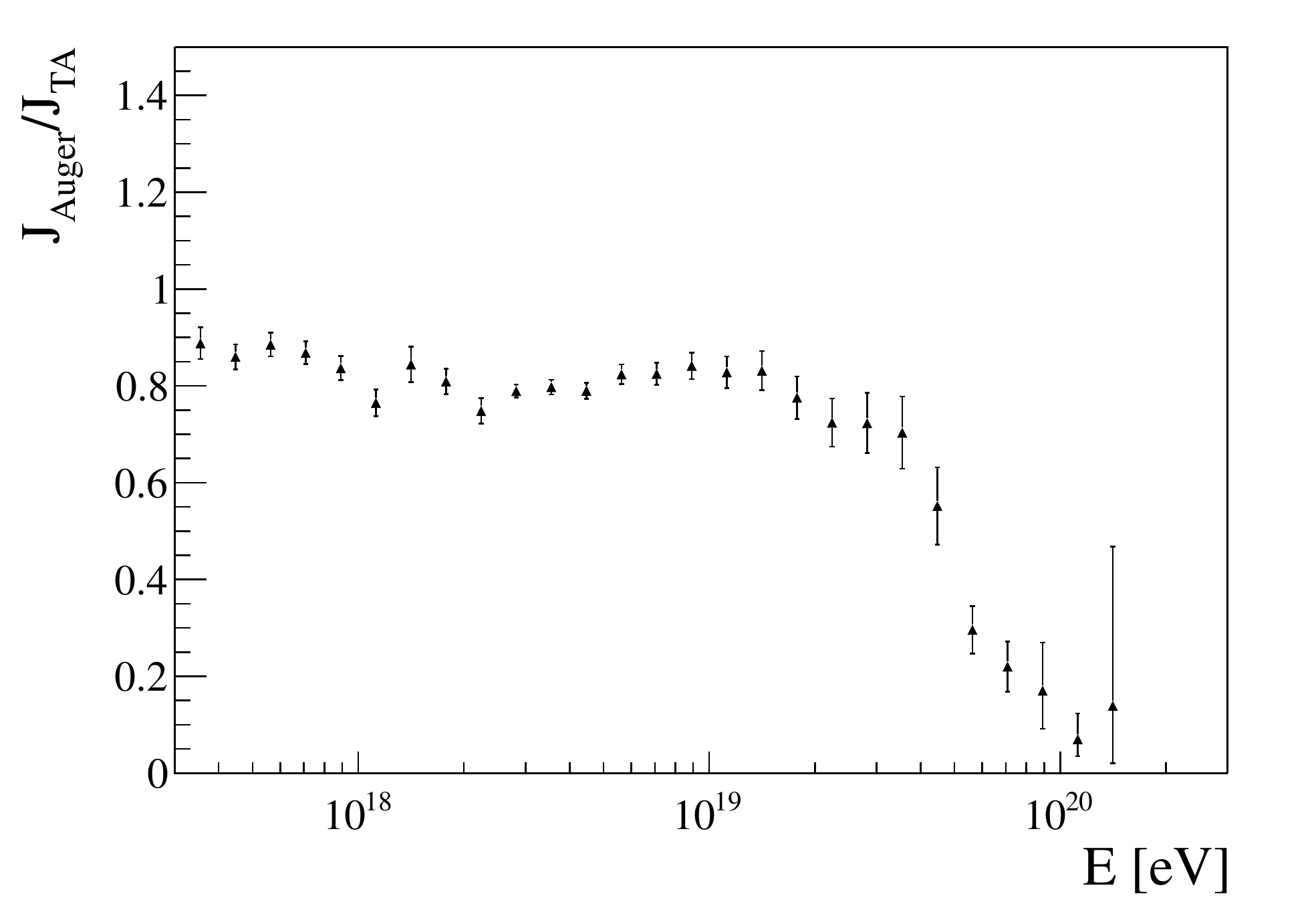}
}
\caption{Left panel: comparison between the TA and Auger combined
  spectra presented at the 34rd International Cosmic Ray Conference
  (ICRC 2015)~\cite{TA-EnSp-Comb,Auger-EnSp}. The TA spectrum is shown
  in the energy range where Auger data are available. The ratio of the
  Auger flux to the TA flux versus energy is plotted in the right
  panel.}
\label{Fig:EnSp_TA_Auger}
\end{figure}

Combined energy spectra of TA and Auger above $3 \times 10^{17}$ eV
are presented in Fig.~\ref{Fig:EnSp_TA_Auger} (left panel).  There is
clearly an overall energy scale difference between the two
measurements, which is emphasized by the multiplication of the two
results by the third power of the energy.  The offset appears to be
constant below the cut-off energy, above which the TA flux becomes
significantly higher than that of Auger.

A more quantitative statement can be made by considering the ratio of
the Auger and TA fluxes, shown in the right panel of
Fig.~\ref{Fig:EnSp_TA_Auger}. Below $\sim2\times10^{19}$~eV, the Auger
flux is $\sim$20\% lower than the TA flux and the difference between
the two measurements becomes large for $E > 2\times 10^{19}$~eV.  It
should be noted that below $2\times 10^{19}$~eV, the two spectra agree
within the systematic uncertainties of the two experiments: a shift in
the energy scale of less than 20\% (a negative energy shift for TA or
a positive energy shift for Auger) would bring the two measurements to
an agreement.  This shift is well within the uncertainties described
in Sec. \ref{SubSec:SystUnc}, and it can be attributed to the
different models of the fluorescence yield and/or the {\it invisible}
energy correction used by the two collaborations (see
Sec.~\ref{Sec:EnergyScaleComparison}).

Another way to address the differences between the two measurements is
to compare the fitting parameters of the functional forms that
describe the shapes of the spectra (see
Sec.~\ref{Sec:SpectrumMeasurement}).  The energy of the {\it ankle}
and the spectral indices below ($\gamma_1$) and above ($\gamma_2$) the
{\it ankle} presented in Table~\ref{Table:FitSpectrum_TA} and
\ref{Table:FitSpectrum_Auger} can be compared directly. As expected,
they are in good agreement. In the region of the cut-off, on the other
hand, the comparison is more difficult, since the parameters that
define the two functional forms have different meanings.  However, an
unambiguous comparison can be made using the parameter suggested
in~\cite{Berezinsky1} that defines the position of the observed
cutoff. This is the energy $E_{1/2}$, at which the integral spectrum
drops by a factor of two below that which would be expected in the
absence of the cutoff. $E_{1/2}$ has been calculated by both
collaborations.  For TA, $E_{1/2} = 60 \pm 7$~EeV (statistical error
only)~\cite{TA-EnSp-Comb} and for Auger, $E_{1/2} = 24.7 \pm 0.1
^{+8.2}_{-3.4}$~EeV~\cite{Auger-EnSp} (statistical and systematic
error).  The two values of $E_{1/2}$ are significantly different, even
after taking into account the systematic uncertainties in the energy
scales of the two experiments.

The difference between the TA and Auger spectra in the region of the
cut-off is very intriguing.  Because the TA experiment is in the
Northern hemisphere and Auger is in the Southern hemisphere and the
two experiments look at different parts of the sky, this could be a
signature of anisotropy of the arrival directions of the ultra-high
energy cosmic rays. Moreover the highest energies are the most
promising for the identification of the sources of cosmic rays since
the deflections of the trajectories of the primaries in the galactic
and extra-galactic magnetic fields are minimized. However the
measurement of the spectrum at the cut-off is affected by large
uncertainties.  In addition to the poor statistics, the analysis is
complicated by the steepness of the flux: large spectral index
amplifies the uncertainty of the energy scale and it increases the
unfolding corrections required to take into account the bin-to-bin
migrations due to the finite energy resolution. A continuous and
increasing effort is being made by the two collaborations at
establishing a better control of these effects and evaluation of the
systematic uncertainties.

\section{Discussion}
\label{Sec:Discussion}

The TA and Auger collaborations have developed analyses to constrain
the astrophysical models using measurements of the energy spectrum.
Observed features in the UHECR spectrum can reveal astrophysical
mechanisms of production and propagation of the UHECRs.  Moreover,
thanks to the unprecedented statistics accumulated by the two
experiments, the collaborations have started studying the energy
spectrum in different regions of the sky. This represents a big step
forward in the cosmic ray field: combined analyses of anisotropies of
the arrival directions of cosmic rays, using high statistics whole-sky
data, and the features in the energy spectrum can significantly
improve our understanding of the nature of the UHECRs.

\subsection{Fitting Energy Spectrum to Astrophysical Models}
\label{Sec:AstrophysicalInterpretations-Models}

The basic assumption of the models developed by the TA and Auger is
that UHECRs are accelerated at the astrophysical sources (the {\it
  bottom-up} models).  In fact, most of the so-called {\it top-down}
models, in which the primaries are generated by the decay of the super
heavy dark matter, or topological defects, or exotic particles have
been excluded by strong upper limits on ultra-high energy photons and
neutrino fluxes~\cite{Auger-icrc15-HL,TA-PhotonLimit}.

The basic approach developed by TA and Auger to interpret the UHECR
spectrum consists of assuming a distribution of identical sources, a
mass composition and an energy spectrum at the sources. Then the
spectrum at Earth is simulated taking into account the interactions of
the primaries with the cosmic radiation (CMB, infrared, optical and
ultraviolet) and the magnetic fields encountered during their
path. The models are characterized by the parameters whose values are
determined from the fit to the experimental data. 

\begin{figure}[h]
\centerline{
\includegraphics[width=1.0\textwidth]{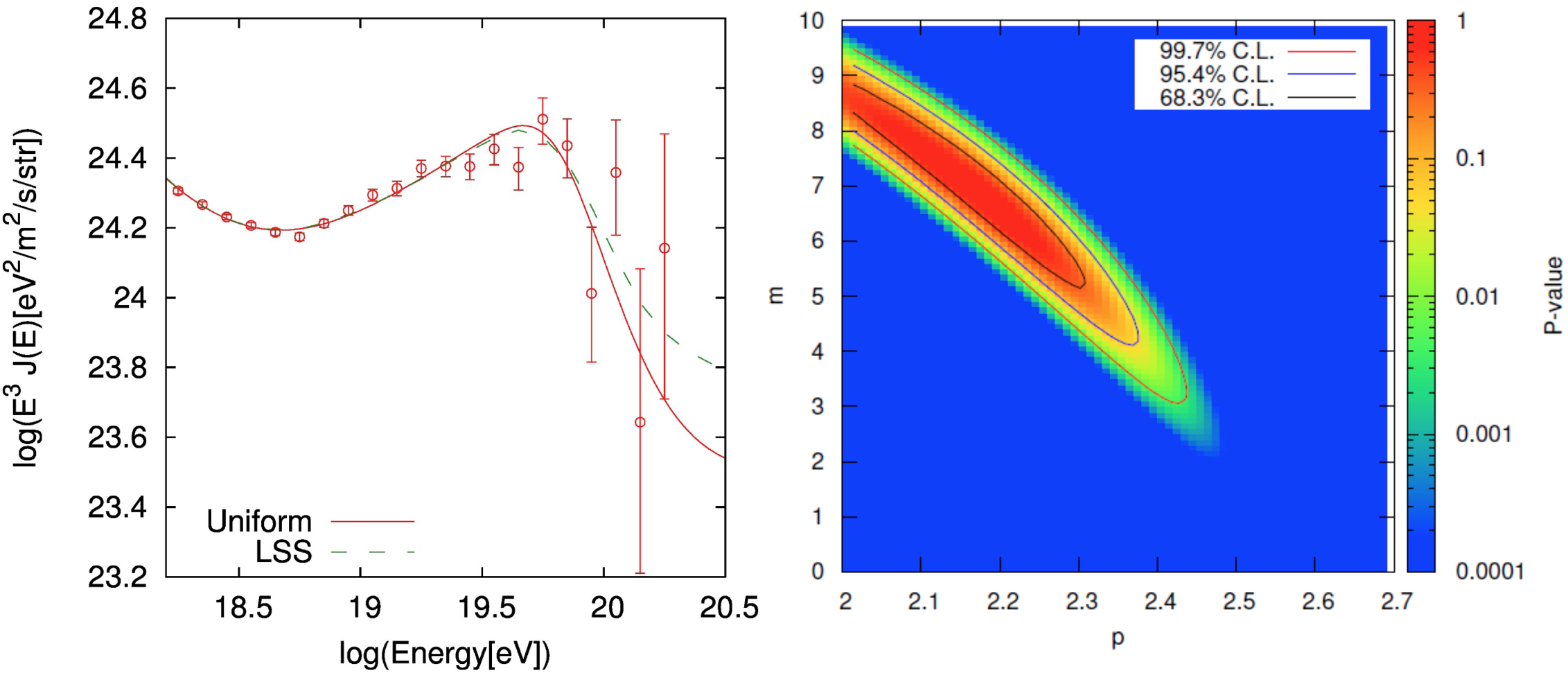}
}
\caption{Interpretation of the TA SD spectrum with astrophysical
  models~\cite{TA-EnSp-Interpretation}, in terms of the cosmic ray
  spectral index at the sources $p$ and the cosmological evolution
  parameter $m$.}
\label{Fig:TA-EnSp-Interpretation}
\end{figure}

In the TA model~\cite{TA-EnSp-Interpretation} the sources are
distributed either uniformly or according to the large-scale structure
(LSS) described by the distribution of the galaxies from the Two
Micron All Sky Survey~\cite{TwoMASS}. Only proton primaries are
simulated.  This composition assumption is justified by measurements
of the mean $X_{\rm max}$ made with the TA FD~\cite{TA-mass}.  The
spectrum of cosmic rays at the sources is parametrized using $\alpha
E^{-p} (1+z)^{3+m}$, where $z$ is the redshift and the parameter $m$
describes the cosmological evolution of the source density (for $m =
0$ the source density is constant per comoving volume).  The maximum
energy at which the primaries are accelerated is fixed at $10^{21}$
eV, well above the energy of the GZK effect.

The results of the TA analysis are shown in
Fig.~\ref{Fig:TA-EnSp-Interpretation}. The model that fits the SD
spectrum~\cite{TA-EnSp-Comb} well is shown using solid and dashed
lines, for uniform and LSS density distribution of the sources,
respectively. The confidence regions of the model parameters are shown
in the right figure. The fitted parameters are $p \approx 2.2$ and $m
\approx 7$~\cite{TA-EnSp-Interpretation}. The latter indicates a very
strong evolution of the sources. The conclusion of the analysis is
that the TA spectrum is well described by the interaction of the
protons with the CMB: the GZK cut-off~\cite{Greisen,ZK}, due to the
photo-pion production, and the {\it ankle} due to the
electron-positron pair production~\cite{Berezinsky1}.

\begin{figure}[h]
\centerline{
\includegraphics[width=1.0\textwidth]{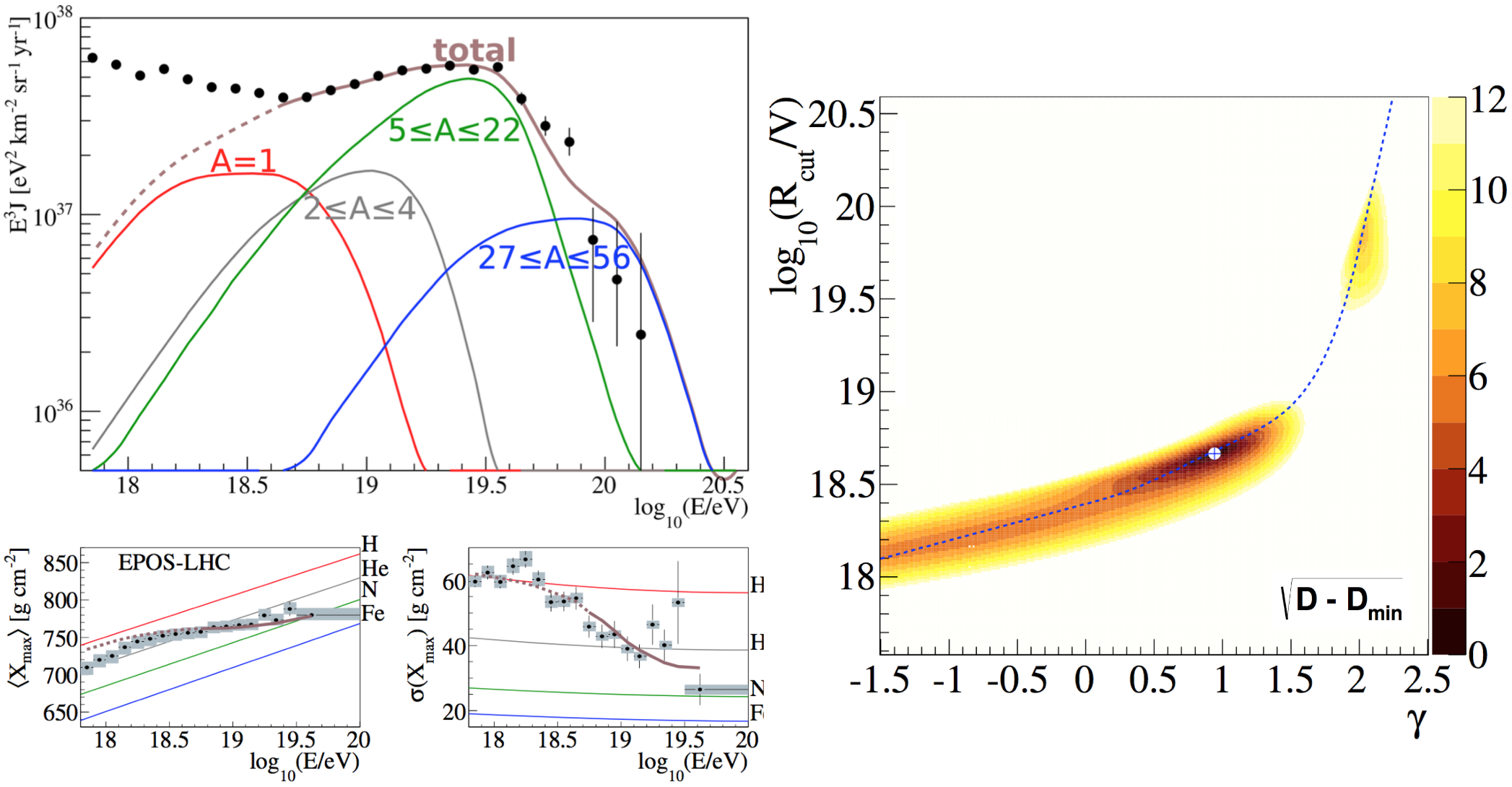} 
}
\caption{Interpretation of the Auger spectrum with astrophysical models~\cite{Auger-EnSp-Interpretation} where 
$X_{\rm max}$ distribution is predicted assuming the EPOS-LHC UHECR air shower interaction model.}
\label{Fig:Auger-EnSp-Interpretation}
\end{figure}
 
In Auger model~\cite{Auger-EnSp-Interpretation,Auger-EnSp-Interpretation-paper},
the UHECR mass composition is not fixed, but is fitted to the Auger
$X_{\rm max}$ data~\cite{Auger-mass}, simultaneously with the fit to
the UHECR energy spectrum~\cite{Auger-EnSp}.  The sources have an isotropic
distribution in a comoving volume. The nuclei are accelerated with a
rigidity-dependent mechanism up to the maximum energy $E_{\rm max} = Z
R_{\rm cut}$ ($Z$ is the charge of the nuclei and $R_{\rm cut}$ is a
free parameter of the model). The spectrum of the sources is
parametrized with $\alpha E^{-\gamma}$.

The results of the analysis are presented in
Fig.~\ref{Fig:Auger-EnSp-Interpretation}. The model that fits the
measured spectrum and the mean and the standard deviation of $X_{\rm
  max}$ best is shown using solid lines in the left panel.  The model
describes the measurements at energies above the {\it ankle}. 
The deviance (equivalent to a $\chi^2$ per degree of freedom)  
as a function of the fitted parameters is 
shown in the right figure. The absolute minimum corresponds to a very hard
injection spectrum ($\gamma \lesssim 1$) and a low maximum
acceleration energy, which is below the energy of the GZK cut-off.
This suggests that the observed break of the spectrum is mainly due to
a cut-off at the sources rather than to the effects of
propagation. There is another less significant minimum at $\gamma
\approx 2$. In this case, the value of $R_{\rm cut}$ is larger and the
propagation effects contribute to the break in the spectrum.

The TA and Auger analyses lead to different conclusions. This is due
to the difference in energies at which the cut-off is observed and to
the different primary mass composition assumptions in the models. In
TA, the primaries are protons, while in Auger, the composition is
mixed and has a trend with energy toward heavier elements in the
suppression region.  It is worth mentioning that the Auger and TA
measurements of $X_{\rm max}$ agree within the systematic
uncertainties~\cite{TA-Auger-mass} but the inferred mass composition
results are different because different hadronic interaction models
and Monte Carlo codes have been used to interpret the data in the two
experiments. Moreover, the sensitivity of the experiments to the mass
composition measurements in the suppression region is strongly limited
by the reduced FD duty cycle, a limitation that the Auger
Collaboration plans to overcome with an upgrade of the SD detector as
described in Sec~\ref{Sec:Outlook}.

\subsection{Study of the Declination Dependence of the Energy Spectrum}
\label{Sec:AstrophysicalInterpretations-DeclinationBands}

\begin{figure}[h]
\centerline{
\includegraphics[width=0.6\textwidth]{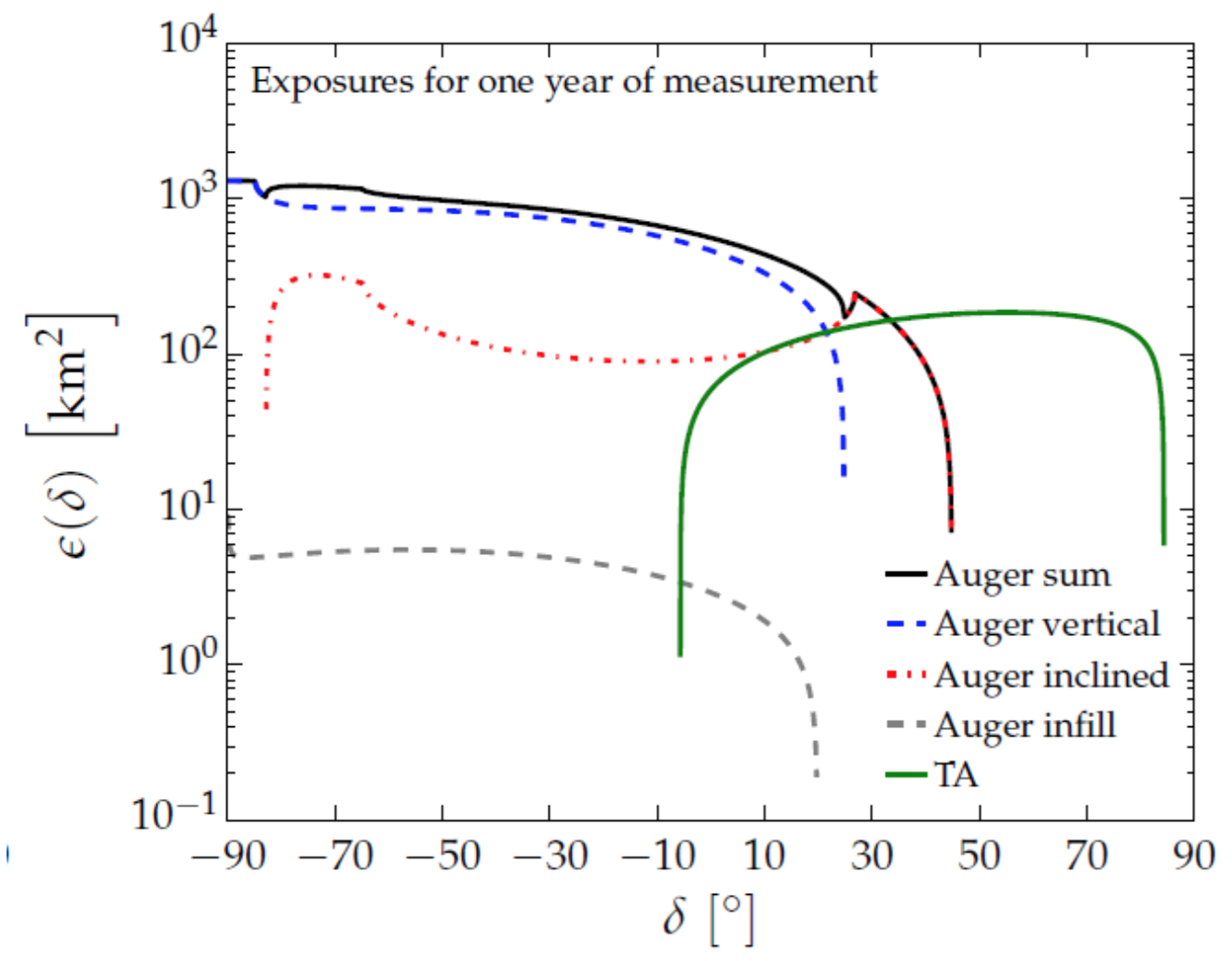}
}
\caption{Exposure as a function of the declination (also called
  directional exposure) for TA and Auger SDs~\cite{Auger-EnSp}. For
  Auger, the exposure is shown for the 750 m array (infill) and for
  the showers detected by the 1500 m array at zenith angles below
  (vertical) and above (inclined) $60^\circ$. The zenith angle range
  for TA is limited to $45^\circ$.}
\label{Fig:Exposure_vs_declination_TA_Auger}
\end{figure}

The TA and Auger collaborations have started studying the energy
spectrum in different declination bands. The exposures of the two SD
detectors versus declination, for one year of data
taking, are shown in
Fig.~\ref{Fig:Exposure_vs_declination_TA_Auger}~\cite{Auger-EnSp}.
For TA, the exposure refers to the events detected by the SD with
zenith angles $\theta$ below $45^{\circ}$. For Auger, the exposures
are for the 750 m ($\theta <55^{\circ}$) and 1500 m ($\theta
<60^{\circ}$ and $>60^{\circ}$) arrays. The Auger exposure obtained by
adding the three contributions is also shown.

The study of the spectrum in different declination bands became
possible due to the large statistics accumulated by the two
experiments.  The study is motivated by the recent indications of
anisotropy of the arrival direction of cosmic rays. The TA
collaboration has found an excess of events of $E > 5.7 \times
10^{19}$~eV in the so called {\it hot spot}, an angular region of
radius $20^\circ$ in the direction of ($\alpha = 148.4^\circ, \delta =
44.5^\circ$ - right ascension and declination), near the Ursa
Major~\cite{TA-HotSpot-publ,TA-Anisotropy}.  The Auger collaboration
reported an indication of a dipole amplitude in right ascension for
the events of energies above $8 \times 10^{18}$ eV, which corresponds
to a reconstructed dipole with $(\alpha,\delta) = (95^\circ \pm
13^\circ , -39^\circ \pm 13^\circ)$.  Also, Auger has found another,
less significant dipole amplitude, at the lower
energies~\cite{Auger-LargeScale}.

\begin{figure}[h]
\centerline{
\includegraphics[width=0.48\textwidth]{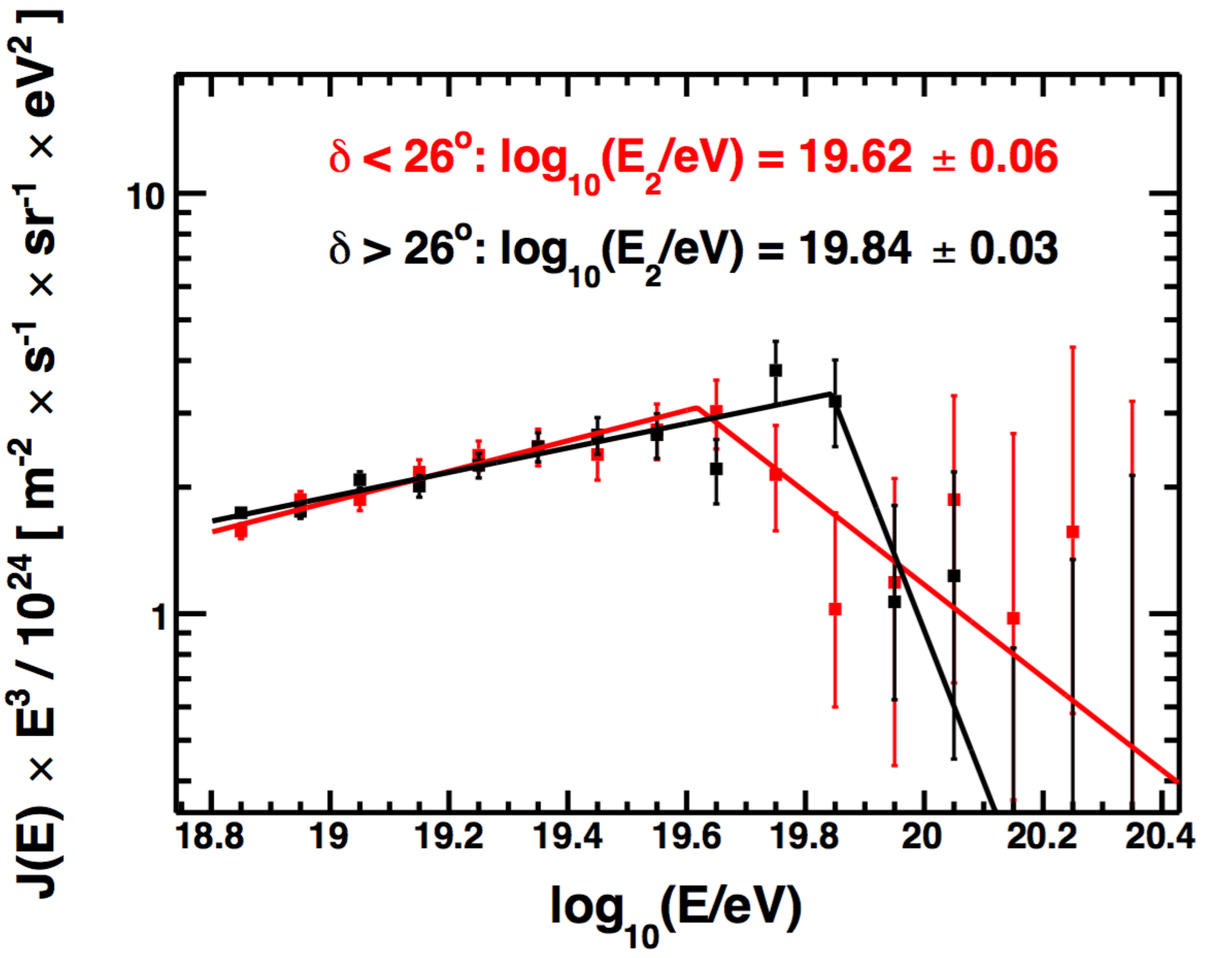}
\includegraphics[width=0.5\textwidth]{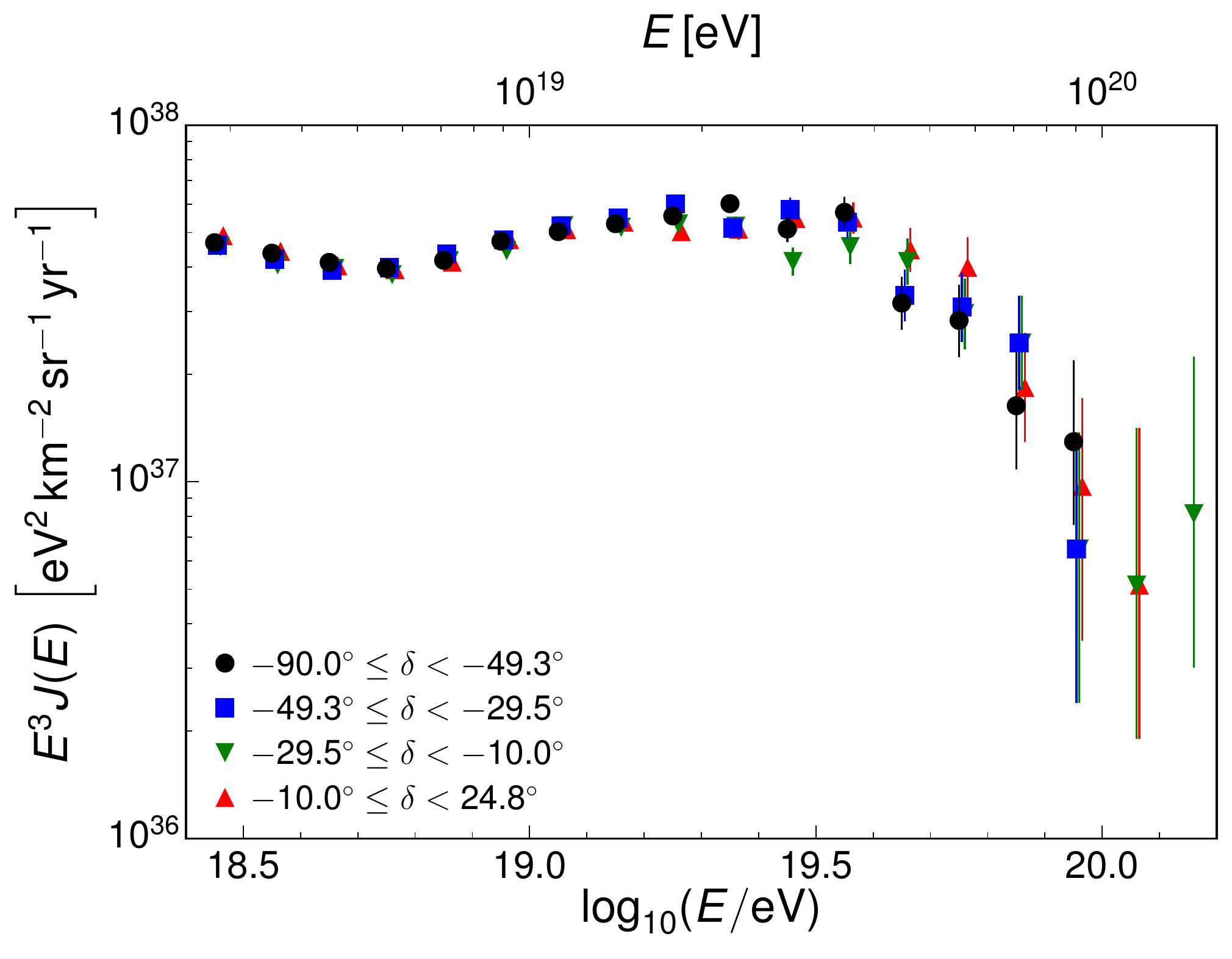}
}
\caption{The preliminary results of the energy spectra of TA~\cite{WG2}
  (left) and Auger~\cite{Auger-EnSp} (right) in different declination
  bands.}
\label{Fig:EnSp_TA_Auger_declination}
\end{figure}

The TA collaboration has measured the SD energy spectrum in two
declination bands, $\delta > 26^\circ$ and $\delta < 26^\circ$.  For
this analysis, TA events with zenith angle ($\theta$) up to $55^\circ$
have been selected. In comparison to the standard SD spectrum
calculation, which is done using events with $\theta < 45^\circ$, this
analysis allows to increase the statistics and to lower the minimum
declination of the events from about $-6^\circ$ to
$-16^\circ$. However it requires an higher energy threshold at
$10^{19}$eV, which is above the {\it ankle}. It has been shown that
the two TA spectra calculations are fully consistent above $10^{19}$
eV~\cite{WG3}.

The declination dependence of the TA spectrum using 6 years of
data~\cite{WG2} is shown in the left panel of
Fig.~\ref{Fig:EnSp_TA_Auger_declination}.  Solid lines represent the
(fitted) power laws with one breaking point. The definition of the
so-called {\it second break point energy} ($E_2$) is equivalent to
$E_{\rm break}$ of Table~\ref{Table:FitSpectrum_TA}.  The
corresponding values are: $E_{\rm break} = (69 \pm 5)$ EeV, for
$\delta > 26^\circ$, and $E_{\rm break} = (42 \pm 6)$ EeV, for $\delta
< 26^\circ$.  Even if the sensitivity of the analysis is low due to
the limited statistics, it is interesting to note that the tension
with Auger data (which observes the suppression at a significantly
lower energy - see Sec.~\ref{Sec:SpectrumComparison}) persists in the
band at larger values of declinations, $\delta > 26^\circ$, while at
the lower declinations, $\delta < 26^\circ$, where the TA and Auger
fields of view partially overlap, the experiments see very similar
energies of the suppression.  

The Auger collaboration has measured the energy spectrum in four
declination bands with an exposure of about 42500/4 km$^2$ sr yr
each~\cite{Auger-EnSp}. The results are presented in the right panel
of Fig.~\ref{Fig:EnSp_TA_Auger_declination}.  There is no significant
declination dependence of the flux. It has been demonstrated that the
small differences between the fluxes are consistent with the
expectation from the dipole anisotropy~\cite{Auger-LargeScale}. The
analysis is limited to the declinations up to $+24.8^\circ$ since it uses
only the events detected by the 1500 m array with zenith angles $<
60^\circ$.

A systematic study of the difference of the spectra measured by the
two experiments in the same declination band is of crucial importance,
since it will help to understand whether the differences between the
spectra addressed in Sec.~\ref{Sec:SpectrumComparison} have been
caused by the systematic uncertainties of the experiments or these
differences are due to an anisotropy signal.  It is worth noting that,
even if the spectra are compared in a declination band accessible by
the two experiments, such analysis would not allow to arrive to a
definitive conclusion if the shapes of the directional exposures in
the common declination band are not similar, because the spectra would
be affected by a potential anisotropy signal in different ways. As
shown in Fig.~\ref{Fig:Exposure_vs_declination_TA_Auger}, this is the
case of the comparison of the TA spectrum with the Auger one obtained
with the vertical events ($\theta < 60^\circ$). In fact, the two
directional exposures have an opposite trend, increasing function of
the declination for TA and decreasing for Auger.

At the time of writing this paper, the Auger collaboration has not
presented the declination dependence of the energy spectrum obtained
using the inclined events ($\theta > 60^\circ$). We remark the
importance of this measurement, since in the common declination band,
the directional exposure for the Auger inclined events is of a similar
shape to that of TA. Moreover, the comparison could be extended to
higher declinations, up to $44.8^\circ$, whereas the vertical event
Auger analysis goes only up to $24.8^\circ$ degrees.

At the 2016 Conference on Ultra-High Energy Cosmic Rays, Kyoto (Japan)
the two collaborations have presented a new and promising analysis
method~\cite{WG3}, proposed by the members of the working group, aimed
at combining the results of the anisotropy searches within the TA and
Auger~\cite{WG-Olivier}. It consists of comparing the results of an
alternative flux estimate, obtained by counting the numbers of events
in the energy bins and weighting them by the inverse of the
directional exposure. The resulting flux does not depend on the shape
of the directional exposure and therefore, it should be same for TA
and Auger. If a difference is found, it is to be ascribed to the
experimental effects, and it should be consistent with the systematic
uncertainties~\ref{SubSec:SystUnc}.

The analysis presented in~\cite{WG3} is still preliminary. However, it
has marked the road that should be followed to understand the
differences in the measurements of the energy spectrum at the highest
energies.  It is worth to note that the application of this method
requires a very good control of the systematic uncertainties. This
alternative flux estimate should be consistent with the standard flux
calculation if the arrival directions of cosmic rays are distributed
isotropically, and this is possible only if the systematic
uncertainties on the event reconstruction and the exposure
calculations are well understood. The TA collaboration has shown that
the two flux estimation methods are consistent in the declination band
accessible by Auger with vertical events ($\delta < 24.8^\circ$) and
in the full declination band $-16^\circ < \delta < 90^\circ$ (which
includes the {\it hot spot}).

\section{Conclusions and Outlook}
\label{Sec:Outlook}

The Telescope Array and the Pierre Auger Observatory are the two
largest cosmic ray detectors built so far.  Their large exposures have
allowed an observation of the suppression of the flux of cosmic rays
at the very high energy with unprecedented statistics and precision.

Both experiments combine the measurements of a surface array with the
fluorescence detector telescopes. The hybrid system allows to measure
the cosmic rays with an almost calorimetric energy estimation, which
is less sensitive to the large and unknown uncertainties due to
limited knowledge in the hadronic models, that are extrapolated well
beyond the energies attainable in laboratory experiments. Having a
precise estimate of the energy scale is of crucial importance for the
measurement of the energy spectrum. In fact, the uncertainty in the
energy estimation ($\Delta E/E$), when propagated to the energy
spectrum ($J$), is amplified by the power index $(\gamma)$ with which
the flux falls off with energy ($\Delta J/J \approx \gamma ~ \Delta
E/E$). 

The TA and Auger measure the cosmic rays in the northern and southern
hemispheres, respectively. At energies below the suppression, the
fluxes are expected to be the same because of the high level of
isotropy in the arrival directions of the cosmic
rays~\cite{TA-icrc15-HL,Auger-icrc15-HL}. A good control of the
systematic uncertainties of the energy scale of the two experiments is
demonstrated by a remarkable agreement attained in the determination
of the {\it ankle} at about $5 \times 10^{18}$ eV. The energy of the
{\it ankle} measured by TA is only +8\% larger than the one measured
by Auger (see Tables~\ref{Table:FitSpectrum_TA}
and~\ref{Table:FitSpectrum_Auger}), which is roughly in agreement with
the 20\% difference in the flux normalization below the cut-off shown
in Fig.~\ref{Fig:EnSp_TA_Auger}. The difference in the {\it ankle}
positions is fully consistent with the uncertainties in the energy
scales quoted by the two experiments (21\% and 14\% for TA and Auger,
respectively) and, it is expected to be reduced if the two
collaborations adopt the same model for the fluorescence yield and for
the {\it invisible} energy correction (see
Sec.~\ref{Sec:EnergyScaleComparison}). 

Despite the good agreement in the region of the {\it ankle} and even
at the lower energies, the TA and Auger spectra differ significantly
in the region of the suppression (see Fig.~\ref{Fig:EnSp_TA_Auger}).
As discussed in Sec.~\ref{Sec:SpectrumComparison}, this discrepancy
can be also quantified comparing the values of the $E_{1/2}$
parameter~\cite{Berezinsky1} that describes the position of the
cut-off. The values reported by the two collaborations differ by a
factor 2.5, which is well beyond the systematic uncertainties of the
energy determination. 

Understanding the difference between the two spectra in the region of
the cut-off is one of the major issues in the study of the UHECRs. At
these extreme energies the deflections of the trajectories of the
primaries in the galactic and extra-galactic magnetic fields are
minimized, allowing the source identification, and therefore the
spectra at Earth detected in the two different hemispheres could be
different. The two collaborations have started studying their spectra
in different declination bands.  For TA, these studies are very
relevant because of the {\it hot spot} near the Ursa Major
constellation~\cite{TA-HotSpot-publ,TA-Anisotropy}.  As shown in
Sec.~\ref{Sec:AstrophysicalInterpretations-DeclinationBands}, these
studies have a great potential but are currently limited by the
statistics. 

Another important finding of these studies is that the declination
range of the exposures of the two experiments partially overlap.  This
offers the possibility of making a comparison of the spectra in the
same region of the sky~\cite{WG1,WG2,WG3}. Any discrepancy found would
be indicative of an experimental effect that's due to the systematic
uncertainties.  One should note that the spectrum steepness in the
energy region of the suppression amplifies the uncertainties in the
energy scale and the event bin-to-bin migration that is due to the
finite energy resolution. These effects, in addition to the limited
statistics, make the measurement of the flux at the energies of the
suppression very challenging.

The features of the energy spectrum at very high energies are
sensitive to the production and the propagation of the UHECRs and have
been used to constrain astrophysical models. As shown in
Sec.~\ref{Sec:AstrophysicalInterpretations-Models}, the TA spectrum is
well fitted by a model in which the primaries are protons (hypothesis
consistent with the TA FD measurement of the mean $X_{\rm
  max}$~\cite{TA-mass}) and therefore the {\it ankle} is explained by
the proton interactions with the CMB via electron-positron pair
production~\cite{Berezinsky1} and the cut-off is explained by the GZK
effect~\cite{Greisen,ZK}.  The Auger interpretation of their energy
spectrum is more complicated. The inclusion of the trend toward
heavier nuclei at the highest energies inferred from the FD
measurements~\cite{Auger-mass} leads to a scenario in which the
observed break of the spectrum is not due to the effects of propagation. In
this model the nuclei are accelerated by a rigidity-dependent
mechanism with a cut-off that is observed in the spectrum measured at
Earth.

The studies presented by the TA and Auger demonstrate that the
knowledge of the chemical composition plays an important role in the
interpretation of the features of the energy spectrum. The results on
$<X_{\rm max}>$ of the two experiments are
consistent~\cite{TA-Auger-mass}, but the inferred mass composition
answers are different because the two collaborations have assumed
different hadronic interaction models and used different Monte Carlo
procedures.  Extrapolation of the hadronic models beyond the energies
attainable by accelerator physics is one of the major issues in
understanding the air showers produced by the UHECRs. The shower
development is mainly influenced by the particle production in the
forward region, where the accelerator data are available only for
energies up to a few hundreds of GeV~\cite{Engel-HadInt}. A big
improvement in this field will be possible by building a fixed target
experiment using the beam of the LHC collider.

The two collaborations will be taking data in the next years and are
working on improving their detectors. The TA collaboration will
quadruple the area of the SD array to approximately the current size
of Auger, which is $~3000$ km$^2$. This extension is called the
TA$\times$4~\cite{TAx4} and it will be realized by adding 500 surface
detectors using 2.08 km spacing.  The aim is to improve the
measurement of the cosmic rays beyond the suppression energy, as well
as the sensitivity to the {\it hot spot} and other astrophysical
sources.  Also, two additional FD stations will be constructed to
overlook the new SD array and to improve the composition studies at
the highest energies.

The Auger collaboration will upgrade the SD array by mounting
scintillator detectors on the top of each WCD station.  The upgrade of
the Pierre Auger Observatory is called AugerPrime~\cite{AugerPrime}.
The combined analysis of the signal of the two detectors will allow to
extract the muonic shower component and to extend the composition
sensitivity of the detector into the flux suppression region, where
the FD measurements are limited by the duty cycle. This will allow to
improve the understanding of the origin of the cut-off and to select
light primaries for the anisotropy studies.

Even if the primary scope of TA and Auger is to study cosmic rays at
the highest energies, an effort has been made with the TALE and Infill
detectors to lower the minimum detectable shower energy threshold.
The TALE FD energy spectrum has made it possible to observe the {\it
  low energy ankle} and the {\it second knee}. A similar result could
be obtained with the HEAT telescopes of Auger. Building surface arrays
of closer spacing is feasible for large collaborations such as TA and
Auger and it would allow to extend the measurements down to the
energies of the {\it knee}.

The next decade will offer many opportunities to understand the origin
of the UHECRs. TA$\times$4 and Auger will view the full sky with a
total collection area of 6000 km$^2$. The two collaborations are
working together on combining their measurements.  The declination
band accessible by the two experiments is instrumental in achieving a
better understanding of the systematic uncertainties and the
differences in the energy scales.  This will allow us to measure the
energy spectrum from the {\it knee} up to the suppression and beyond
in the entire sky with an unprecedented statistics and precision,
which in turn will allows us to measure the energy spectra of cosmic
rays in different declination bands or sky patches. So far, anisotropy
studies using small (a few degree) or intermediate angular scales were
carried out independently from the energy spectrum studies, although
there were several studies of the energy dependencies of anisotropies.
We emphasize here that the energy spectrum, the number of cosmic ray
particles per time in a unit area from a given direction in a given
energy range is, by definition, a function of the direction. The
measurement of the full-sky energy spectrum by the future Auger and TA
will make a crucial contribution to identifying the sources of ultra-
high energy cosmic rays.

\begin{figure}[h]
\centerline{
\includegraphics[width=1.0\textwidth]{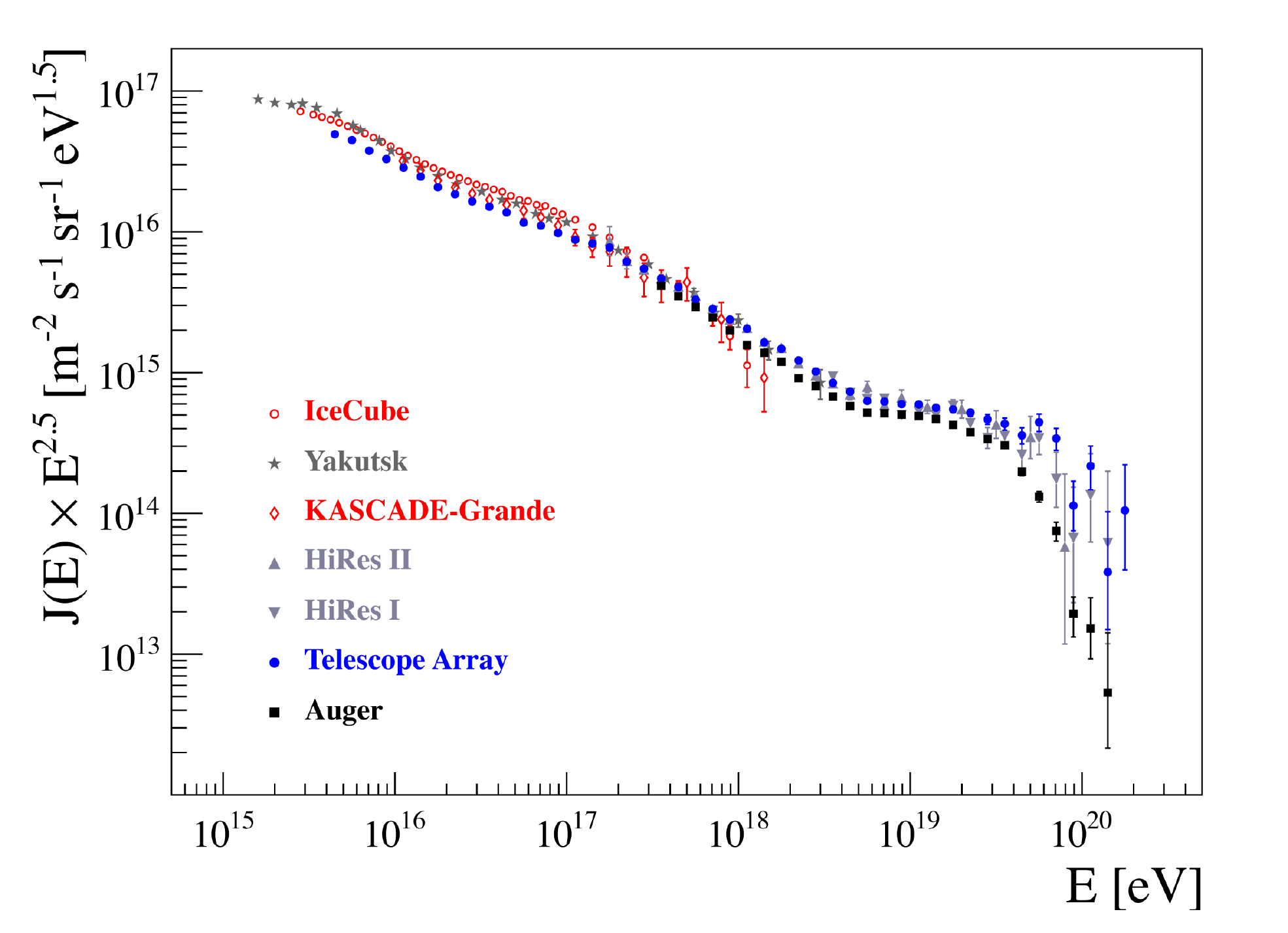}
}
\caption{Energy spectra measured by IceCube~\cite{EnSp-IceTop}, Yakutsk~\cite{EnSp-Yakutsk},  KASCADE-Grande~\cite{EnSp-KASCADE}, HiRes I and HiRes II~\cite{EnSp-Hires2008}, 
Telescope Array~\cite{TA-EnSp-Comb} and Auger~\cite{Auger-EnSp}.
}
\label{Fig:AllSpectra}
\end{figure}
We conclude this review with a compilation of recent experimental data 
on the energy spectrum presented in Fig.~\ref{Fig:AllSpectra}.

\end{document}